\documentclass[12pt]{article}
\usepackage[dvips]{epsfig}
\usepackage{latexsym}
\usepackage{graphicx}
\usepackage{psfrag}
\usepackage[bf,footnotesize]{caption}	
\begin{document}
\setlength{\unitlength}{1mm}


\newcommand{\ket}[1] {\mbox{$ \vert #1 \rangle $}}
\newcommand{\bra}[1] {\mbox{$ \langle #1 \vert $}}
\def\vac{\ket{0}} 
\def\thermal{\ket{\beta}}
\def\bvac{\bra{0}}
\def\bthermal{\bra{\beta}}
\newcommand{\ave}[1] {\mbox{$ \langle #1 \rangle $}}
\newcommand{\vacave}[1] {\mbox{$ \bvac #1 \vac $}}
\newcommand{\thermalave}[1] {\mbox{$ \bthermal #1 \thermal $}}
\newcommand{\scal}[2]{\mbox{$ \langle #1 \vert #2 \rangle $}}
\newcommand{\expect}[3] {\mbox{$ \bra{#1} #2 \ket{#3} $}}
\def\aa{\hat{a}}\def\AA{\hat{A}}
\def\bb{\hat{b}}\def\BB{\hat{B}}
\def\a{\tilde{a}}\def\A{\tilde{A}}
\def\b{\tilde{b}}\def\B{\tilde{B}}


\def\p{\prime}
\def\t{\tau}
\def\ga{\gamma}\def\Ga{\Gamma}
\def\om{\omega}\def\Om{\Omega}
\def\omp{\om^\p}\def\Omp{\Om^\p}
\def\la{\lambda}\def\lap{\lambda^\p}
\def\mup{\mu^\p}\def\lp{l^\p}
\def\kp{k^\p}\def\sig{\sigma} 
\def\al{\alpha}\def\alb{\bar\alpha}
\def\bt{\beta}\def\btb{\bar\beta}
\def\e{\epsilon}
\def\psip{\stackrel{.}{\psi}}\def\fp{\stackrel{.}{f}}
\def\VB{{\bar V}}\def\UB{{\bar U}}
\def\vb{{\bar v}}\def\ub{{\bar u}}
\def\lab{{\bar \la}}
\def\ffi{\varphi}
\def\scryp{{\cal J}^+}
\def\scrym{{\cal J}^-}
\def\scrypL{{\cal J}^+_L}
\def\scrymR{{\cal J}^-_R}
\def\scrypR{{\cal J}^+_R}
\def\scrymL{{\cal J}^-_L}
\def\Pig{{\mathbf \Pi}}
\def\div{\partial_V}
\def\divp{\partial_{V'}}
\def\divb{\partial_{\VB}}
\def\di{\partial}
\newcommand{\didi}[1]{\raise 0.1mm \hbox{$\stackrel{\leftrightarrow}{\di_{#1}}$}}


\def\TVV{T_{VV}}\def\TUU{T_{UU}}
\def\TVVI{T_{VV}^I}\def\TVVII{T_{VV}^{II}}
\def\TVVB{T_{{\VB}{\VB}}}
\def\Tvv{T_{vv}}
\def\Tvvb{T_{{\vb}{\vb}}}
\def\re{\mbox{Re}}
\def\im{\mbox{Im}}
\def\S{{\mathbf S}}
\def\T{{\mathbf T}}
\def\1{{\mathbf 1}}
\def\Ss{\mbox{$\hat S$}}

\def\disp{\displaystyle}
\def\bitem{\begin{itemize}}
\def\eitem{\end{itemize}}
\def\bes{\begin{description}}
\def\es{\end{description}}
\newcommand{\be} {\begin{equation}}
\newcommand{\ee} {\end{equation}}
\newcommand{\ba} {\begin{eqnarray}}
\newcommand{\ea} {\end{eqnarray}}

\def\cf{{\it cf}~}
\def\ie{{\it i.e.}~}
\def\etc{{\it etc}...}
\def\where{\mbox{where} \; \;}
\def\whereas{\mbox{whereas} \; \;}
\def\with{\mbox{with} \; \;}
\def\for{\mbox{for} \; \;}
\def\and{\mbox{and} \; \;}
\def\eg{{\it e.g.}~}

\def\nn{\nonumber \\}
\newcommand{\reff}[1]{Eq.(\ref{#1})}


\def\g{ g_+}
\def\sgn{\mbox{sgn}}
\def\half{{1 \over 2}}
\newcommand{\inv}[1]{\frac{1}{#1}}

\def\inte{\int_{-\infty}^{+\infty}}
\def\into{\int_{0}^{\infty}}

\newcommand\Ie[1]{\inte \! d #1 \;}
\newcommand\Io[1]{\into \! d #1 \;}


\def\M{Minkowski}
\def\sg{\sqrt{-g}}
\def\sga{\sqrt{-\gamma}}
\def\gmn{g_{\mu\nu}}
\def\gnm{g^{\mu\nu}}
\def\gmn{g_{\mu\nu}}
\def\Rmn{R_{\mu\nu}}
\def\Rnm{R^{\mu\nu}}
\def\Gmn{G_{\mu\nu}}
\def\Gnm{G^{\mu\nu}}
\def\Tmn{T_{\mu\nu}}
\def\Tnm{T^{\mu\nu}}
\def\gnm{g^{\mu\nu}}
\def\Rij{R_{ij}}
\def\Rji{R^{ij}}
\def\Kij{K_{ij}}
\def\Kji{K^{ij}}
\def\gij{g_{ij}}
\def\gji{g^{ij}}
\def\piij{\pi_{ij}}
\def\piji{\pi^{ij}}
\def\gaij{\gamma_{ij}}
\def\gaji{\gamma^{ij}}
\def\H{{\cal H}}
\def\Hi{{\cal H}_i}
\def\Hii{{\cal H}^i}
\newcommand{\G}[3]{\Gamma^{#1}_{#2 \: #3}}
\def\A{{\cal A}_{BH}}

\overfullrule=0pt \def\sqr#1#2{{\vcenter{\vbox{\hrule height.#2pt
          \hbox{\vrule width.#2pt height#1pt \kern#1pt
           \vrule width.#2pt}
           \hrule height.#2pt}}}}
\def\lrpartial{\mathrel{\partial\kern-.75em\raise1.75ex\hbox{$\leftrightarrow$}}}

\newcounter{subequation}[equation] \makeatletter
\expandafter\let\expandafter\reset@font\csname reset@font\endcsname
\newenvironment{subeqnarray}
  {\arraycolsep1pt
    \def\@eqnnum\stepcounter##1{\stepcounter{subequation}{\reset@font\rm
      (\theequation\alph{subequation})}}\eqnarray}%
  {\endeqnarray\stepcounter{equation}}
\makeatother



\vskip 1. truecm
\vskip 1. truecm
\centerline{\Large\bf{Uniformly accelerated mirrors}}
\vskip .2 truecm
\centerline{\large\bf{Part II : Quantum correlations}}
\vskip 1. truecm
\vskip 1. truecm

\centerline{{\bf N. Obadia}\footnote{e-mail: obadia@celfi.phys.univ-tours.fr}
 and {\bf
R. Parentani}\footnote{e-mail: parenta@celfi.phys.univ-tours.fr}}
\vskip 5 truemm\vskip 5 truemm
\centerline{Laboratoire de Math\'ematiques et Physique Th\'eorique,
CNRS-UMR 6083}
\centerline{Parc de Grandmont, 37200 Tours, France.}
\vskip 10 truemm

\vskip 1.5 truecm

\vskip 1.5 truecm
\centerline{{\bf Abstract }}
\vskip 3 truemm
\noindent
We study the correlations between the particles emitted by a moving mirror.
To this end, we first analyze $\ave{T_{\mu\nu}(x) T_{\al\bt}(x')}$,
the two-point function of the stress tensor of the radiation field.
In this we generalize the work undertaken by Carlitz and Willey.
To further analyze how the vacuum correlations on $\scrym$ are scattered 
by the mirror and redistributed among the produced pairs of particles,
we use a more powerful approach 
based on the value of $T_{\mu\nu}$ which is conditional 
to the detection of a given particle on $\scryp$.
We apply both methods to the fluxes emitted by a uniformly accelerated mirror.
This case is particularly interesting because of its strong interferences
which lead to a vanishing flux, 
and because of its divergences
which are due to the infinite blue shift effects associated with the horizons.
Using the conditional value of $T_{\mu\nu}$,
we reveal the existence of correlations between created particles
and their partners in a domain where the mean fluxes and the
two-point function vanish. This demonstrates that the scattering
by an accelerated mirror leads to a steady conversion
of vacuum fluctuations into pairs of quanta. 
Finally, we study the scattering by two uniformly accelerated mirrors
which follow symmetrical trajectories (\ie which possess the same horizons).
When using the Davies-Fulling model, the Bogoliubov coefficients encoding pair creation
vanish because of perfectly destructive interferences. 
When using regularized amplitudes, 
these interferences are inevitably lost thereby giving rise to pair creation.

\vfill
\newpage


\section*{Introduction}

It is now well understood that the scattering of a quantum radiation field
by a non-inertial mirror leads to the production of pairs of 
particles\cite{DaviesFulling,BirrelDavies,PhysRep}.
However, up to now, most studies have been restricted to the analysis 
of mean quantities such as the expectation value of the stress tensor 
$\ave{T_{\mu\nu}(x)}$, \ie the one-point function.
This analysis is very restrictive in that most of the information 
concerning the correlations among particles is ignored.
In particular, $\ave{T_{\mu\nu}(x)}$ cannot be used to 
identify the relationships between the particles
and their partners.

In this paper, it is our intention to go beyond the mean field approach.
To this end, we first study the 
(connected part of the)
two-point function $\ave{T_{\mu\nu}(x)T_{\al\bt}(x')}_c$.
In this, we complete the analysis undertaken by Carlitz and Willey 
\cite{CarlitzWilley} and Wilczek \cite{Wilczek}, see also \cite{WuFord,Hu1}.
Our motivations are the following. 
First, since $T_{\mu\nu}$ is the source of gravity,
if one wishes to go beyond the semi-classical treatment,
\ie Einstein equations driven by the mean $\ave{T_{\mu\nu}}$,
it is imperative to gain some experience concerning the two-point function
since it governs the metric fluctuations 
about the mean background geometry \cite{York,Casher,Verdaguer,Parentani}.
Secondly, we wish to relate the analysis of $\ave{T_{\mu\nu}(x)T_{\al\bt}(x')}_c$
to an alternative approach\cite{Aharonov,MaPaelectro,MaPa,PhysRep} of
correlations which was used to 
reveal the space time distribution of the correlations
among charged pairs produced in a constant electric
field and among Hawking quanta emerging from a black hole.
This method is based on the value of $T_{\mu\nu}$ which is conditional to the detection 
of a specific quantum (or specific quanta) on $\scryp$. 
We shall show that the two approaches are closely related
and that the second one is more powerful to identify the correlations between
the particles and their partners.
Finally, the quantum correlations within the fluxes emitted by
a mirror constitute an interesting subject {\em per se}.

In this respect, it is particularly interesting to study the correlations in
the fluxes emitted by a uniformly accelerated mirror.
Indeed, these fluxes possess, on one hand, 
strong interferences which 
lead to a vanishing mean flux and, on the other hand, 
very high frequencies associated with
the diverging blue shift effects encountered when the mirror
enters or leaves space-time. 
In order to tame this singular behavior, one needs to abandon
the original Davies-Fulling model \cite{DaviesFulling}
and use a self-interacting model described by an action \cite{Recmir,OPa,OPa2}.
In this paper, we shall compare the two-point functions 
computed with the Davies-Fulling model and this self-interacting model.

Because of the strong interferences
in the case of uniform acceleration, we shall see that 
the analysis of the two-point function is not sufficient
to properly isolate the correlations among the produced particles. 
To complete the analysis, we therefore use 
the conditional value of $T_{\mu \nu}$. 
By an appropriate choice of the detected quantum on  $\scryp$,
we unravel correlations among the two members in a produced pair
even in domains where the mean flux and the two-point function vanish. 
These correlations show that the scattering by a uniformly accelerated mirror
leads to a steady conversion of vacuum fluctuations into pairs of particles, 
something which could not be seen from the expressions of the mean flux and the 
two-point function which both vanish.
Another nice property of this alternative approach is that
the wave packet of the detected particle can be chosen in such a way that
the former regularization of the scattering amplitudes is no longer necessary.
We hope that this double and complementary analysis of observables 
in the presence of very high frequencies 
can lead to a better understanding of the ``trans-Planckian'' physics, 
\ie the fact that Hawking 
radiation\cite{Jacobson91+Jacobson00,tHooft,Unruh95,BMPS95,MaPa,Parentani}, 
and cosmological density fluctuations\cite{MBrand+N,NP}
arise from ultra-high energy configurations.

Finally, to illustrate the necessity of using regular scattering amplitudes, 
we study the scattering by two 
uniformly accelerated mirrors which follow symmetrical trajectories 
(\ie which possess the same horizons).
When using the Davies-Fulling model,
the Bogoliubov coefficients governing pair creation
identically vanish. This vanishing follows
from perfectly destructive interferences between the two 
mirrors, a phenomenon related to what Gerlach\cite{Gerlach} called a perfect interferometer 
and which was also found when considering the fluxes emitted by two
accelerating black holes\cite{Yi,MaPaYi}.
When using regulated amplitudes, 
we show that these interferences  are inevitably lost
and that the total energy emitted is 
the sum of the energy emitted by each mirror.
It thus appears that the perfect interferences are an artifact due to 
the oversimplification of the description of the scattering.
A similar conclusion can 
be reached when taking into account recoil effects
\cite{Recmir,OPa4}.
This further legitimizes the use of regulated scattering amplitudes.

We have organized the paper as follows.
In Sec. $1$, we recall the basic properties of the self-interacting model.
Sec. $2$ is devoted to the study of the two-point correlation function.
In Sec. $3$ we compute the conditional value of $T_{\mu\nu}$ 
and in Sec. $4$ we study 
the scattering by two uniformly accelerated mirrors.

\section{The Lagrangian model}

In \cite{OPa2}, our aim was to obtain 
regular expressions for the fluxes and
the energy emitted by a uniformly accelerated mirror.
To this end, the scattering of the scalar field $\Phi$ by the mirror was described by 
a self-interacting model based on an action. 

The action density is localized on the mirror trajectory 
$x^\mu_{cl}(\t)$ where $\t$ is the proper time.
To preserve the linearity of the scattering,
the density is a quadratic form of the field $\Phi$.
Since the field is massless,
IR divergences appear in the transition amplitudes.
To get rid of these problems, it is sufficient to use a density which contains
two time derivatives.
The interacting Lagrangian we shall use is
\ba \label{Hint}
L_{int} 
&=& - g_0 \int d\tau g(\t) \int \! d^2x \: \delta^2(x^\mu-x^\mu_{cl}(\t)) \;
\frac{dx^\mu_{cl}}{d\t} \frac{dx^\nu_{cl}}{d\t}
(\di_\mu \Phi^\dagger \di_\nu \Phi + \di_\mu \Phi \di_\nu \Phi^\dagger) \\
&=& -g_0 \int d\tau g(\t) 
\left( \di_\t \Phi^\dagger \di_\t \Phi
+ \di_\t \Phi \di_\t \Phi^\dagger \right) \nonumber \ .
\ea
Here, $g_0$ is the coupling constant.
The real function $g(\t)$ controls the time dependence of the interaction.
When the interaction lasts $2T$, its normalization is given by $\int d\t g(\t) = 2T$.
The two terms in the parentheses imply that $L_{int}$ is charge-less.
Hence the transition amplitudes will be invariant under charge conjugation.

We work in the interacting picture. Therefore,
the charged field evolves freely, \ie according to the 
d'Alembert equation, 
\ba
\Box \Phi(t,z) = 4 \di_U \di_V \Phi(U,V) = 0 \ .
\ea
Since the field is massless,
it is useful to use the light-like coordinates 
$U,V=t \mp z$.
The free field $\Phi$ can be decomposed as
\ba \label{field}
\Phi(U,V) = \into \! \frac{d\om}{\sqrt{4\pi \om}} 
\left(
a^{U}_{\om} e^{-i\om U}
+ a^{V}_{\om} e^{-i\om V}
+ b^{U \: \dagger}_{\om} e^{i\om U}
+ b^{V \: \dagger}_{\om} e^{i\om V}
\right)  \ .
\ea   
The annihilation and creation operators 
of left and right-moving particles (and anti-particles)
are constant and obey the usual commutation relations 
\ba
[a^i_\om,a^{j \: \dagger}_{\omp}] = \delta^{ij} \delta(\om-\omp)
\; \; , \; \; 
[b^i_\om,b^{j \: \dagger}_{\omp}] = \delta^{ij} \delta(\om-\omp)
\ ,
\ea
where the indices $i,j$ stand for $U$ and $V$. 
All other commutators vanish.
In the interacting picture, 
the states evolve through the action of the time-ordered operator 
$Te^{iL_{int}}$.
When the initial state is vacuum, 
the state on $\scryp$ is given, up to second order in $g_0$, by
\ba \label{evoluatedvac}
T e^{i L_{int}} \vac 
&=& \vac 
+ i L_{int}  \vac + \frac{(iL_{int} )^2}{2} \vac
+ \ket{D} \ .
\ea
The ket $\ket{D}$ contains 
terms arising from time-ordering. 
These terms do not contribute to the total energy emitted (see \cite{OPa}).
Hence we drop $\ket{D}$ from now on. 

When working in the vacuum, to obtain the mean flux $\ave{T_{\mu\nu}}$
and the two-point function $\ave{T_{\mu\nu}T_{\al\bt}}$
to order $g_0^2$, 
it is sufficient to develop the scattering 
amplitudes to first order in $g_0$.
To this order, the amplitude describing the scattering 
of an incoming quantum of frequency $\omp$ 
to an outgoing of frequency $\om$ is given by
\ba \label{Bogo1}
A^{ij \; *}_{\om \omp}
\equiv
\vacave{a^i_\om \:  (1+iL_{int}) \: a^{j \; \dagger}_{\omp}}_c \ ,
\ea
where the subscript $\langle \rangle_c$ means that 
only the connected graphs are kept.
Similarly, the spontaneous pair production amplitude reads
\ba \label{Bogo2}
B^{ij}_{\om \omp}
\equiv
\vacave{ a^{i}_\om b^{j}_{\omp} \: iL_{int} \: } \ .
\ea

Both non-local and local objects are easily obtained
in terms $A$ and $B$. 
For instance,  to order $g_0^2$,
the mean number of spontaneously created 
left-moving particles of frequency $\om$ is given by
\be
\ave{N^V_\om} \label{N}
\equiv \vacave{L_{int} \: 
a^{V \: \dagger}_\om a^V_\om \: {L_{int}}}
= \Io{\omp} 
\left( \left| B^{VV}_{\om \omp} \right|^2 
+ \left| B^{VU}_{\om \omp} \right|^2 \right) \ . 
\ee
Then, the (subtracted) integrated energy is, as usual,
\ba
\ave{H^V} &=& 2 \Io{\om} \om \ave{N^V_\om} \ . \label{H}
\ea
The factor of $2$ stands for particles + antiparticles,
which contribute equally.
One can also compute the local flux of energy.
The corresponding operator is
$\TVV = \di_V \Phi^\dagger \di_V \Phi + \di_V \Phi \di_V \Phi^\dagger$. 
Its vacuum expectation value is given by 
\ba
\ave{\TVV(V)} &\equiv& \label{TVV1} 
\vacave{e^{-iL_{int}} \: \TVV \: e^{iL_{int}}}_c  - \vacave{\TVV} 
\nonumber \\ 
\label{TVVI}
&=& 2 \sum_{j=U,V} \int \! \Io{\om} \! \! d\omp 
\frac{\sqrt{\om \omp}}{2 \pi} 
e^{-i(\omp-\om)V} 
\left( \Io{k} B^{Vj \: *}_{\om k} B^{Vj}_{\omp k} \right) \nn
&& - 2 \re \Bigg\{
\sum_{j=U,V} \int \! \Io{\om} \! \! d\omp 
\frac{\sqrt{\om \omp}}{2 \pi}  
e^{-i(\omp+\om)V} 
\left( \Io{k} A^{Vj \: *}_{\om k} B^{Vj}_{\omp k} \right)
\Bigg\} \ .
\ea
We have subtracted the average value of $\TVV$ in the vacuum 
in order to remove the zero point energy.
When integrated over all $V$,
the first term of \reff{TVVI} determines the (positive) energy 
$\ave{H^V}$ of \reff{H}.
The second term clearly integrates to $0$.

The above model can easily be related to the original Davies-Fulling one
\cite{DaviesFulling,BirrelDavies},
where the field obeys
\ba \label{EqDF}
\Box \Phi(U,V) = 0 \; \; \and \Phi(V,V_{cl}(U)) = 0 \ ,
\ea
where $V_{cl}(U)$ is the mirror trajectory expressed in null coordinates.
On the right of the trajectory, 
the Bogoliubov coefficients are given by the overlaps
between  initial $V$ modes of frequency $\om$ defined on $\scrymR$
and $out$-modes of frequency $\omp$ defined on $\scrypR$:
\ba
\al_{\om \omp}^{*} \label{aUV}
&\equiv& ( \ffi_{\om}^{U , out} , \ffi_{\omp}^{V , in} )
= - \Ie{U} \frac{e^{i\om U}}{\sqrt{4\pi \om}}  i \didi{U} 
\frac{e^{-i\omp V_{cl}(U)}}{\sqrt{4\pi \omp}} \ , \\
\bt_{\om \omp}^{*} \label{bUV}
&\equiv& ( \ffi_{\om}^{U , out \: *} , \ffi_{\omp}^{V , in} )
= - \Ie{U} \frac{e^{-i\om U}}{\sqrt{4\pi \om}}  i \didi{U} 
\frac{e^{-i\omp V_{cl}(U)}}{\sqrt{4\pi \omp}} \ .
\ea
Two properties are worth remembering.
In the Davies-Fulling model, when starting from vacuum,
the flux of energy emitted by the mirror
is given by  \reff{TVVI} with
$A^{VU}_{\om \omp}$ and $B^{VU}_{\om \omp}$ 
respectively replaced by $\al_{\om \omp}$ and $\bt_{\om \omp}$, 
and $A^{VV}_{\om \omp},B^{VV}_{\om \omp} $ sent to zero.
Secondly, $A^{UV}_{\om \omp} = i g_0 \al_{\om \omp}$ 
and $B^{UV}_{\om \omp} = - i g_0 \bt_{\om \omp}$ for all
trajectories $V=V_{cl}(U)$
when using $\Phi^\dagger i \didi{\tau} \Phi$ instead of 
$\di_{\tau}\Phi^\dagger \di_{\tau}\Phi + \di_{\tau}\Phi \di_{\tau}\Phi^\dagger$ 
in \reff{Hint} and when putting $g(\t) = 1$,  see \cite{OPa} for more details.

These two properties guarantee that $\ave{\TVV(V)}$ behave similarly
whether one uses the Lagrangian or the Davies-Fulling model
to describe the scattering. It is only for trajectories which lead to singular fluxes
when using the Davies-Fulling model that the two descriptions can significantly
differ because the coupling function $g(\tau)$ 
in \reff{Hint} can be chosen so as to obtain regular expressions.

\section{The two-point correlation function}

To analyze the quantum correlations 
among the particles emitted by the mirror,
we first study the two-point function of $T_{\mu\nu}$.
Given the Lagrangian defined in \reff{Hint},
when a $U$ quantum is detected on ${\scrypR}$
(the right hand side of ${\scryp}$, see Fig.$1$),
its partner can be either a $U$ or a $V$ quantum,
emitted respectively toward ${\scrypR}$ or ${\scrypL}$. 

In this Section, for reasons of simplicity,
we mainly focus on $U/V$ correlations and study 
the two-point function $\ave{\TUU\TVV}_c$.
Indeed, in the absence of the mirror, these correlations vanish.
Hence, if $\ave{\TUU\TVV}_c \neq 0$, it results from the scattering
and not from pre-existing correlations which exist in the vacuum, see Sec. $2.1$.
This is not the case for $\ave{\TVV\TVV}_c$ which originates both 
from the scattering as well as from pre-existing correlations. 
Moreover, since these two channels interfere, 
the expressions are much more complicated.

\subsection{Initial correlations on $\scrym$, before the scattering}

On $\scrymR$, when the trajectory does not enter space through it,
the field is unscattered.
Therefore, when working in the vacuum,
the two-point function is given by
\ba
C_{vac}(V,V') &\equiv& \vacave{\TVV(U=-\infty,V) \: \TVV(U'=-\infty,V')}_c \nn
&=& (2 \di_V \di_{V'} W_{vac}(V,V') )^2
= \inv{4\pi^2} \frac{1}{(V-V'-i\e)^4} \ , \label{Cvac}
\ea
where the subscript $c$ means that only connected graphs are kept
and where $W_{vac}(V,V')$ is the $V$-part of the vacuum Wightman function 
\ba \label{Wvac}
W_{vac}(V,V') \equiv \vacave{\Phi(U=-\infty,V)\: \Phi^\dagger(U'=-\infty,V')} 
= - \inv{4\pi} \ln \left( V-V'-i\e \right) \ .
\ea

\reff{Cvac} is valid for the Davies-Fulling model and for the Lagrangian model.
It applies both for an inertial and for a uniformly accelerated mirror in $L$,
since the past null infinity $\scrymR$ lies on the past of the mirror, see Fig.$2$.
Had the mirror entered space through  $\scrymR$, a prescription
should have been adopted  to define the in-vacuum on $\scrymR$
in the presence of a mirror, see Sec. $1.2$ in \cite{OPa2}.

\subsection{Correlations between $\scrym$ and $\scryp$}

When the field is initially in vacuum,
the correlations between $\scrym$ and $\scryp$ are governed by 
the (connected) two-point function
\ba \label{defCUV-+}
C^{+/-}(U,V') \equiv \vacave{T_{UU}(U,V=+\infty) \: T_{VV}(U'=-\infty,V')}_c \ .
\ea
In this expression, written in the Heisenberg picture, 
only $\TUU$ is evaluated on $\scryp$. Hence,  in the interacting picture, 
$C^{+/-}(U,V')$ is given by
\ba \label{CUV-+1}
C^{+/-}(U,V')
&\equiv&
\vacave{e^{-iL_{int}} T_{UU}(U) e^{iL_{int}} \: T_{VV}(V')}_c \ ,
\ea
where $T_{UU}$ and is $T_{VV}$ are now expressed in terms of the free field of \reff{field}.
To second order in the coupling constant and 
when neglecting again the $\ket{D}$ term of \reff{evoluatedvac}, 
we get
\ba 
C^{+/-}(U,V')
&=& \vacave{L_{int} T_{UU} L_{int} T_{VV} }_c \nn
&& - \half \vacave{(L_{int} L_{int} T_{UU} + T_{UU} L_{int} L_{int}) T_{VV}}_c
\label{CUV-+2}\ .
\ea
To compute this expression, it is convenient to introduce the functions
\ba \label{FUV}
F(U,V') &\equiv& i \vacave{\di_U \Phi(U) \di_{V'} \Phi^{\dagger}(V') \: L_{int} }  \\
&=& - \int \! \! \Io{\om} \! \! d\omp \frac{\sqrt{\om\omp}}{4\pi} B_{\om\omp}^{UV} 
e^{-i(\om U +\omp V')} \label{FUVtrans}
\ea
and 
\ba\label{GUV}
G(U,V') &\equiv& i \vacave{\di_U \Phi(U) \: L_{int} \: \di_{V'} \Phi^{\dagger}(V')} \\
&=& \int \! \! \Io{\om} \! \! d\omp \frac{\sqrt{\om\omp}}{4\pi} A_{\om\omp}^{UV \: *} 
e^{-i(\om U -\omp V')} \label{GUVtrans} \ .
\ea From 
these equations,
we see that $F$ is expressed in terms of 
the pair production amplitude $B$ whereas
$G$ is a function of the scattering amplitude $A$.
Using Eqs.(\ref{FUV}) and (\ref{GUV}),
one can rewrite the correlation function in the following form
\ba 
C^{+/-}(U,V')
= \left( {F^*(U,V') +  G(U,V')}  \right)^2 \label{CUV-+3} \ .
\ea

Before applying \reff{CUV-+3} to inertial and accelerated trajectories, 
it is interesting to compute \reff{defCUV-+} in the Davies-Fulling model.
In this case, one gets 
\ba 
C^{+/-}_{DF}(U,V') 
&=& 2 \left( \di_U \di_{V'} W^{+/-}_{DF}(U,V') \right)^2 \nn
&=& \inv{4\pi^2} \frac{(dV_{cl}/dU)^2}{(V_{cl}(U)-V'-i\e)^4} \ ,
\label{CUV-+DF}
\ea
where the relevant part of the scattered Wightman function is given by 
\ba 
W^{+/-}_{DF}(U,V') 
&=& \vacave{\Phi(U,V=+\infty)\Phi^\dagger(U'=-\infty,V')} \nn
&=& \inv{4\pi} \ln \left( V_{cl}(U) - V' - i\e \right) \ . \label{WUV-+DF}
\ea

Eqs.(\ref{CUV-+3}) and (\ref{CUV-+DF}) show that the two-point function is not real. 
In this regard, it is worth making the following remarks.
First we note that $C^{+/-}_{DF}$ diverges only when $V' \to V_{cl}(U)$, 
\ie at the classical image point. Similarly, we shall see 
that $C^{+/-}$ of \reff{CUV-+3} also diverges in this limit
but is otherwise finite. The fact that no regularization is needed to
evaluate it follows from the fact that  only connected graphs
have been kept in \reff{CUV-+1}.

Secondly, we note that the imaginary character of  $C^{+/-}_{DF}$ 
arises only from its singular limit. The sign of $\e$ encodes the fact that only
positive frequencies enter in the Wightman function \reff{Wvac}.  
Therefore, on one hand, the limit $\e \to 0$ can be taken when
evaluating the two-point functions when $V' \neq V_{cl}(U)$.
On the other hand however, the $i \e$
prescription must be kept when using $C^{+/-}$
to obtain the correlations of integrated operators, \eg
$\ave{H^U T_{VV}}$ where ${H^U} = \int dU {T_{UU}}$. 
Indeed, the definition of the integral over $U$ 
requires that the limit $\e \to 0$ be taken after having performed
the integral, see \cite{ParRindler} where it is shown that the
same procedure should be used to properly evaluate the
energy in the Rindler vacuum.
In this sense, the two-point functions should be viewed as distributions.


\subsubsection{Inertial mirror}

In the case of an inertial mirror,
the canonical\footnote{Generally, inertial trajectories read
$V_{cl}(U) = |\xi| (U-U_0) + V_0$. They all provide 
the same two-point functions by applying 
$U' = \sqrt{|\xi|} (U-U_0)$, $V' = (V-V_0)/\sqrt{|\xi|}$.} 
trajectory reads $V_{cl}(U)=U$. 
The corresponding space-time diagram is pictured in Fig.$1$.
\begin{figure}[ht] 
\epsfxsize=5.5cm
\centerline{{\epsfbox{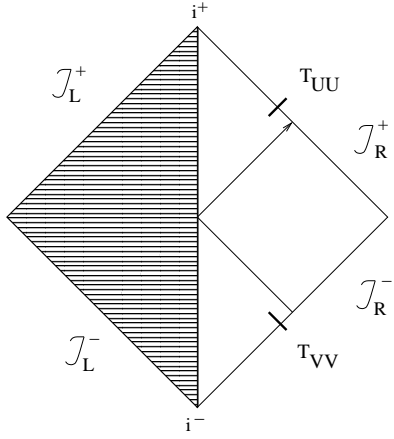}}}
\caption{The Penrose diagram of space-time around an inertial mirror.
Being interested in the correlations between $\scrymR$ and $\scrypR$, 
the dashed region is of no interest.}
\end{figure}
In the Davies-Fulling model, using \reff{CUV-+DF}, the two-point function reads 
\ba
C^{+/-}_{DF,inert}(U,V') = \inv{4\pi^2} \frac{1}{(U-V'-i\e)^4} \label{CUV-+inertDF} \ .
\ea
The small distance divergence comes from the vacuum configurations 
emerging from $\scrymR$ at $V'$ and 
which have been reflected on the mirror at $V'=V_{cl}(U)=U$.
We therefore recover the divergence which existed in the vacuum
on $\scrym$, see \reff{Cvac}.

To compute the corresponding two-point function in the Lagrangian model,
we use \reff{CUV-+3}. Eqs.(\ref{FUV}) and (\ref{GUV}) applied to the inertial trajectory
give
\ba
F_{inert}(U,V') &=& -2ig_0 \Ie{t} g(t) 
\di_U \di_t W_{vac}(U,t) \di_{V'} \di_t W_{vac}(V',t) \nn
&=& \frac{-2ig_0}{(4\pi)^2} \Ie{t} g(t)
\inv{(U-t-i\e)^2}\inv{(V'-t-i\e)^2} \label{FUVinert1} \\
G_{inert}(U,V') &=& -2ig_0 \Ie{t} g(t) 
\di_U \di_t W_{vac}(U,t) \di_t \di_{V'} W_{vac}(t,V') \nn
&=& \frac{-2ig_0}{(4\pi)^2} \Ie{t} g(t)
\inv{(U-t-i\e)^2}\inv{(t-V'-i\e)^2} \label{GUVinert1} \ .
\ea
To obtain local expressions of $F_{inert}$ and $G_{inert}$, 
we express $g(t)$ with its Fourier components 
and perform the integrations over $t$ by the method of residues.
Since Minkowski vacuum contains only positive frequency, 
it is appropriate to decompose $g(t)$ as  $g(t) = \g(t)  + g_-(t)$
where $\g$ contains only positive frequencies:
\ba \label{gtilde}
\g(t) = \Io{\om} g_\om e^{-i\om t} \; \; 
\with g_\om \equiv \inv{2\pi} \Ie{t} g(t) e^{i\om t} \ .
\ea
Then we get 
\ba
F_{inert}(U,V') &=& - \frac{g_0}{4\pi} 
\left[ 
\di_U \left( \frac{\g(U)}{(U-V'-i\e)^2} \right)
+ \di_{V'} \left( \frac{\g(V')}{(V'-U-i\e)^2} \right)
\right] \label{FUVinert2} \\
G_{inert}(U,V') &=& - \frac{g_0}{4\pi} 
\left[
\di_U \left( \frac{\g(U)}{(U-V'-i\e)^2} \right)
- \di_{V'} \left( \frac{g_-(V')}{(V'-U+i\e)^2} \right)
\right] \label{GUVinert2} \ .
\ea

To obtain the value of $C^{+/-}_{inert}(U,V')$ off the
image point $U = U_{cl}(V')=V'$, we can take the limit $\e \rightarrow 0$.
Using \reff{CUV-+3}, we get 
\ba 
C^{+/-}_{inert}(U,V') \vert_{\e = 0} = \frac{g_0^2}{(4\pi)^2}
\left[
\di_U \left( \frac{g(U)}{(U-V')^2} \right)
\right]^2 \label{CUV-+inertg} \ .
\ea
Two remarks should be made. 
On one hand, when $g(t)\equiv 1$, one recovers \reff{CUV-+inertDF} up to two differences.
The first one is the $g_0^2$ pre-factor due to the perturbative expansion.
The second one is the power of the pole (of order $6$ instead of $4$).
This discrepancy arises from the fact that 
we have taken a Lagrangian with two time derivatives (see \reff{Hint}).
Had we taken the current $J=\Phi^\dagger i\didi{\tau} \Phi$
instead of $\di{\tau} \Phi^\dagger \di{\tau} \Phi +  \di{\tau} \Phi \di{\tau} \Phi^\dagger$, 
we would have obtained a fourth order pole in \reff{CUV-+inertg}.
However, in that case, the expressions in Fourier transform,
\ie Eqs.(\ref{FUVtrans}) and (\ref{GUVtrans}), 
would have been divergent in the low-frequency limit.

On the other hand, when $g(t)$ possesses a compact support, 
the correlator identically vanishes, as it should do by causality, 
when the {\it outgoing} configurations (here the $U$ configurations)
lie outside the support of $g(t)$, 
\ie when they cross the trajectory without being scattered by the mirror.
Hence, unlike in the Davies-Fulling model, 
$C^{+/-}_{inert}(U,V')$ is not symmetric in $U, V'$.


\subsubsection{Classical scattering by a uniformly accelerated mirror}
 
We first look at classically reflected configurations,
\ie at configurations which have support for $V'<0$ and $U>0$
when the uniformly accelerated trajectory follows $V_{cl}(U) = - 1/a^2 U$
in the left Rindler wedge, see Fig.$2$.
This situation corresponds to what we just analyzed for an inertial trajectory.
\begin{figure}[ht] 
\epsfxsize=6.5cm
\centerline{{\epsfbox{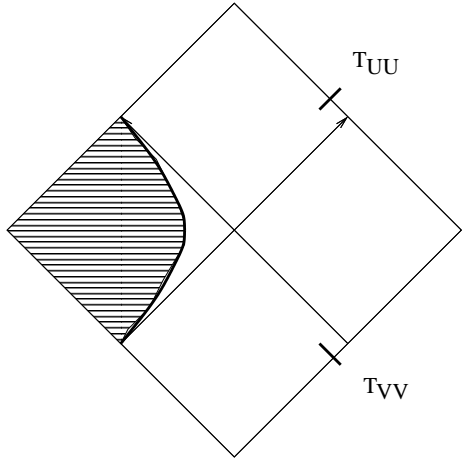}}}
\caption{The Penrose diagram of space-time around a uniformly accelerated mirror
in the left Rindler wedge. As before, we consider only configurations on the right
of the trajectory. }
\end{figure}

In the Davies-Fulling model, the relevant part of the Wightman function,
\reff{WUV-+DF}, is given by
\ba
W^{+/-}_{DF,unif.acc.}(U,V') 
= \inv{4\pi} \Theta(U) \ln \left(- 1/(a^2U)-V'-i\e \right) \ .
\ea 
Then, \reff{CUV-+DF} gives 
\ba
C^{+/-}_{DF,unif.acc.}(U,V') = \inv{4\pi^2} \Theta(U)
\frac{(1/a^2U^2)^2}{(-1/(a^2U)-V'-i\e)^4} \label{CUV-+uaDF1}  \ ,
\ea
up to an ill-defined singular contribution on $U=0$.
Using the Rindler coordinates 
\ba
v_L' = -\inv{a} \ln(-aV') \; \; \and u_L = \inv{a} \ln(aU) \ ,
\ea
the ``Rindler'' correlation function is
\ba
C^{+/-}_{DF,unif.acc.}(u_L,v_L')
&\equiv& \ave{T_{uu}(u_L)T_{vv}(v_L')} \nn
&=& (\frac{dU}{du_L})^2 (\frac{dV'}{dv_L'})^2 \, C^{+/-}_{DF,unif.acc.}(U,V') \nn
&=& \left[ \frac{a^2}{8\pi} \label{CUV-+uaDF2}
\inv{\sinh^2(\frac{a(u_L-v_L'-i\e)}{2})}
\right]^2 \ .
\ea
When comparing this 
expression\footnote{We note that a similar expression has been obtained 
by Carlitz and Willey\cite{CarlitzWilley} in the case of the trajectory 
$- \kappa U= \ln\left( -\kappa V_{cl}(U) \right)$.
When defining $- \kappa v_L' = \ln(-\kappa V')$ for $V'<0$,  
their result reads
\ba \label{CUV-+CW} 
C^{+/-}_{DF,CW}(U,v_L')
= \left[ \frac{\kappa^2}{8\pi} 
\inv{\sinh^2(\frac{\kappa(U-v_L'-i\e)}{2})}
\right]^2 \ .
\ea}  
to that of an inertial trajectory, 
\reff{CUV-+inertDF}, one finds the following correspondence:
\reff{CUV-+uaDF2} is exactly the expression one would have obtained 
for $C^{+/-}$ in the case of an inertial mirror 
in a thermal bath with $T=a/2\pi$, see \cite{Wilczek}.
This is no surprise since the scattering of Rindler modes by an 
accelerated mirror is identical (in fact trivial when using the Davies-Fulling model)
to the scattering of Minkowski modes by an inertial mirror, 
and since the two-point function $W_{vac}$ of \reff{Wvac} is thermal 
when expressed in terms of Rindler coordinates.
We shall now verify that the same correspondence applies 
to the Lagrangian model.

When applying this model to the uniformly accelerated trajectory, 
Eqs.(\ref{FUV}) and (\ref{GUV}) give
\ba
F_{unif.acc.}(U,V') 
&=& \frac{-2ig_0a^4}{(16\pi)^2} e^{a(v_L'-u_L)}\Ie{\t} g(\t) \label{FUVacpoles}
\inv{\sinh^2(\frac{a(u_L-\t-i\e)}{2})} 
\inv{\sinh^2(\frac{a(v_L'-\t-i\e)}{2})} \\
G_{unif.acc.}(U,V') 
&=& \frac{-2ig_0a^4}{(16\pi)^2} e^{a(v_L'-u_L)}\Ie{\t} g(\t) 
\inv{\sinh^2(\frac{a(u_L-\t-i\e)}{2})} \label{GUVacpoles}
\inv{\sinh^2(\frac{a(\t-v_L'-i\e)}{2})} \ .
\ea
As in the inertial case, one can perform the integrals when using the following decomposition
\ba
\inv{\sinh^2(x)} = \sum_{n=-\infty}^{+\infty} \inv{(x-in\pi)^2} \ ,
\ea
and by introducing the function
\ba 
\bar g_+(\t) = \Ie{\la} \frac{g_\la e^{-i\la \t}}{1-e^{-2\pi\la/a}} \; \; 
\with g_\la \equiv \inv{2\pi} \Ie{\t} g(\t) e^{i\la \t} \ ,
\ea
and $\bar g_-(\t) = \bar g_+^*(\t)$.
We get
\ba
F_{unif.acc.}(U,V') = - \frac{g_0}{4\pi} \frac{a^2}{4} e^{a(v_L'-u_L)} 
&&\left[
\di_{u_L} \left( 
\frac{\bar g_+ (u_L)}{\sinh^2(\frac{a}{2}(u_L-v_L'-i\e))} \right) 
\right. \nn
&& \left. +\di_{v_L'} \left( 
\frac{\bar g_+ (v_L')}{\sinh^2(\frac{a}{2}(v_L'-u_L-i\e))} \right) 
\right] \,  
\label{FUVuag}
\ea
\ba
G_{unif.acc.}(U,V') = - \frac{g_0}{4\pi} \frac{a^2}{4} e^{a(v_L'-u_L)}
&&\left[
\di_{u_L} \left( 
\frac{\bar g_+ (u_L)}{\sinh^2(\frac{a}{2}(u_L-v_L'-i\e))} \right)
\right.  \nn
&& \left. -\di_{v_L'} \left( 
\frac{\bar g_- (v_L')}{\sinh^2(\frac{a}{2}(v_L'-u_L+i\e))} \right)
\right] \label{GUVuag}\ .
\ea
Since $\bar g_+ (\t) + \bar g_- (\t) = g(\t) $, 
when taking the limit $\e \rightarrow 0$, 
the correlation function reads
\ba
C^{+/-}_{unif.acc.}(u_L,v_L') = \frac{g_0^2}{(4\pi)^2}
\left[ \frac{a^2}{4} 
\di_{u_L}  \left( \frac{g(u_L)}{\sinh^2(\frac{a}{2}(u_L-v_L'))} \right)
\right]^2 \label{CUV-+uag} \ ,
\ea
as expected from Eq.(\ref{CUV-+inertg}) 
and the correspondence between accelerated systems
described in Rindler coordinates and inertial systems
in a thermal bath.

Moreover, when $g(\t)$ is a constant,
we recover that the only differences 
between the two-point functions obtained in the Davies-Fulling model
and in the Lagrangian model concern
the $g_0^2$ pre-factor and the additional proper time derivative.   
In fact these relations are generic since they directly follow from the fact that,
when $g(\t)$ is a constant,
the scattering amplitudes of Eqs.(\ref{Bogo1}) and (\ref{Bogo2}) are proportional 
to the Bogoliubov coefficients obtained in the Davies-Fulling model,
see Eqs.($33$) in \cite{OPa}. 


\subsubsection{Other correlations on the right of an accelerated mirror}

When the mirror is uniformly accelerated, in addition
to the ``classical'' scattering analyzed above, there exists three other sectors
since both $\scrym$ and  $\scryp$ cover two Rindler patches.

Let us first examine the trivial correlations between $\scrym$ and $\scryp$.
They are obtained for $U<0$ and any $V'$, \ie below the past horizon of the mirror, see Fig.$3$. 
Non surprisingly, these correlations identically vanish.
Indeed, causality tells us that these correlations are equal 
to the (null) correlations between $U$ and $V$ vacuum configurations 
evaluated on $\scrym$
\be
C^{+/-}(U<0,V') 
= C^{-/-}(U<0,V') = 0 \ . 
\label{vanishCUV-+ua}
\ee

\begin{figure}[ht] 
\epsfxsize=13cm
\centerline{{\epsfbox{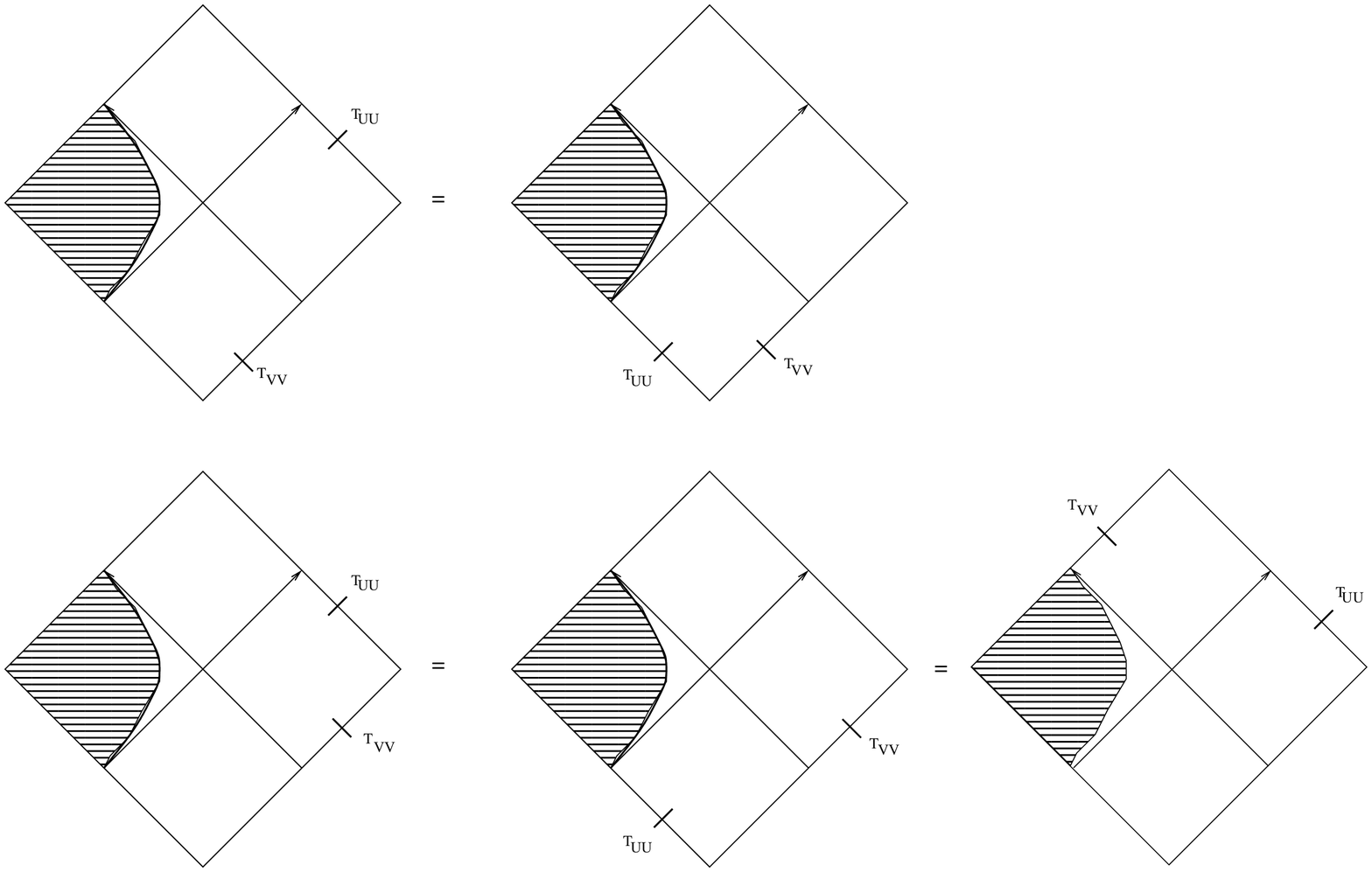}}}
\caption{In these Penrose diagrams, we show the equivalence between 
the correlations between $\scrymR$ and $\scrypR$ with $U<0$ 
and those between $\scrymR$ and $\scrymL$.
Since the latter identically vanish, so do the formers.
This is also the case in the fifth diagram which represents $U/V$
correlations on $\scryp$ when $U <0$, see Sec. $2.3$.}
\end{figure}

The last and most interesting case corresponds to the correlations between $U>0$
and $V'>0$. In the Davies-Fulling model, 
the correlation function is again given by \reff{CUV-+uaDF1}.
Indeed, the Wightman function given in \reff{WUV-+DF} is valid for any $V'$.
Then, given the Rindler coordinate on the other side of the horizon $V=0$,
\ba
v_R' = \inv{a} \ln(aV') \ ,
\ea
one can rewrite the correlation function in the following way
\ba
C^{+/-}_{DF,unif.acc.}(u_L,v_R') = \left[ \frac{a^2}{8\pi}
\inv{\cosh^2\left( \frac{a}{2}(v_R'+u_L)\right)} 
\right]^2 \ .\label{CUV-+uaDF3} 
\ea
This result can be obtained by analytically continuing \reff{CUV-+uaDF1} 
according to $V'\to V' e^{-i\pi}$. It is the $i\e$ prescription which 
specifies that the continuation should be performed in the lower plane.
In terms of Rindler coordinates, it amounts to apply $v_L \to -v_R +i\pi/a$ 
in \reff{CUV-+uaDF2}.
This analytical continuation also applies to the Unruh modes \cite{PhysRep}
and follows from the fact that Minkowski vacuum contains 
positive frequencies only.\footnote{This continuation also applies for
the trajectory $\kappa U= -  \ln( - \kappa V)$ for  $V<0$.
When defining $\kappa v'_R = \ln(\kappa V')$ for $V'>0$,
Carlitz and Willey noticed that 
\ba\label{CUV-+CW2}
C^{+/-}_{DF,CW}(U,v'_R)
= \left[ \frac{\kappa^2}{8\pi} 
\inv{\cosh^2(\frac{\kappa(U+v'_R)}{2})}
\right]^2 
\ea
is the analytical continuation of \reff{CUV-+CW}.}

In the interacting model, since $V'>0$ lies in a disconnected region for the trajectory, 
the field expressed at this point commutes with the Lagrangian.
Hence, one has
\ba
F(U>0,V'>0) 
&=& G(U>0,V'>0) 
\label{feqg} \\
&=& \frac{-2ig_0a^4}{(16\pi)^2} e^{-a(v_R'+u_L)}\Ie{\t} g(\t) 
\inv{\sinh^2(\frac{a(u_L-\t-i\e)}{2})} 
\inv{\cosh^2(\frac{a(v_R'+\t-i\e)}{2})} \ . \nonumber
\ea
Thus, \reff{CUV-+3} gives
\ba
C^{+/-}_{unif.acc.}(U>0,V'>0) = 4 \left( \re [ F(U>0,V'>0) ] \right)^2 \ ,
\ea
which is real, as $C^{+/-}_{DF}$ in \reff{CUV-+uaDF3}.
The corresponding Rindler two-point function is given by 
\ba\label{CUV-+uag2}
C^{+/-}_{unif.acc.}(u_L,v_R') = \frac{g_0^2}{(4\pi)^2}
\left[ \frac{a^2}{4} 
\di_{u_L}  \left( \frac{g(u_L)}{\cosh^2(\frac{a}{2}(u_L+v_R')} \right)
\right]^2 \ .
\ea
It is interesting to notice that \reff{CUV-+uag2} could have been obtained 
in two different ways. 
On one hand, \reff{CUV-+uag2} follows from \reff{CUV-+uaDF3}
by applying the generic relations between the two models.
On the other hand, \reff{CUV-+uag2} could have been obtained from \reff{CUV-+uag}
because the (regularized) scattering amplitudes, see \reff{crossym},
obey crossing symmetry
which follows from the stability of Minkowski vacuum and which allows
to deform the integral over $\omp$ so as to obtain \reff{feqg}.

Since Eqs.(\ref{CUV-+uaDF3}) and (\ref{CUV-+uag2}) 
do not concern the classical reflexion on the mirror, they never diverge.
In fact, they are expressed only in terms of pair creation coefficients
which decrease in the ultraviolet regime like
$B^{VU}_{\om\omp} \sim  e^{-\sqrt{\om\omp}/a}$.
Nevertheless, they are peaked around $u_L+v_R'=0$
with a spread governed by $1/a$. 
This locus corresponds to $V=-V_{cl}(U)$.
This maximum indicates that the configurations which give rise to the ``partner'' 
of an outgoing quantum, found for $U>0$, are vacuum configurations
which are symmetrically distributed on the other side of the horizon,
see Fig.$4$. 

This result is somehow paradoxical in the case
of a uniformly accelerated mirror because 
it indicates that pairs are steadily produced (in terms of the proper time)
in a domain where the mean flux $\ave{T_{UU}}$ vanishes
(Instead, the steady character of \reff{CUV-+CW2} causes no surprise since
the corresponding $\ave{T_{UU}}$ is thermal and constant\cite{CarlitzWilley}).
To clarify the situation, we shall analyze in Sec. $3$ the correlations between 
$U$ and $V$ configurations by using the alternative method
based on the conditional value of $T_{\mu \nu}$. 
Before doing so, it is interesting to 
analyze the correlations between $\scrypR$ and $\scrypL$.

\begin{figure}[ht] 
\epsfxsize=6.5cm 
\centerline{{\epsfbox{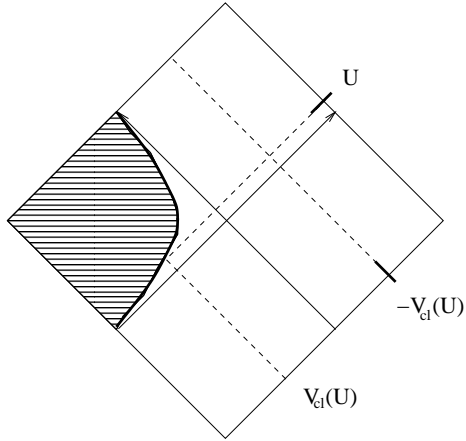}}}
\caption{This diagram illustrates the fact that 
the correlations between $T_{UU}$ and $T_{VV}$ when $U>0$ and $V'>0$
are peaked around $V= -V_{cl}(U)$, that is, on the other side
of the horizon at $V=0$, and symmetrically with respect to the locus
of classical reflection given by $V=V_{cl}(U)$,
see Eqs(\ref{CUV-+uaDF3}) and (\ref{CUV-+uag2}).
The diagram also illustrates the fact that,
by causality, $C^{+/+}$ of \reff{corrCUV-+uag2}
is equal $C^{+/-}$ of  \reff{CUV-+uaDF3} when $U>0$ and $V'>0$.}
\end{figure}

\subsection{The correlations on $\scryp$}

We now look at the $U/V$ correlations on $\scryp$, that is
\ba\label{CUV++0}
C^{+/+}(U,V') \equiv \vacave{T_{UU}(U,V=+\infty)T_{VV}(U'=+\infty,V')}_c \ .
\ea
In the interacting picture, the two-point function reads
\ba 
C^{+/+}(U,V') 
= \vacave{e^{-iL_{int}} T_{UU}(U) T_{VV}(V') e^{iL_{int}}}_c \ .\label{CUV++1}
\ea
To second order in $g_0$, we get
\ba 
C^{+/+}(U,V')
&=& \label{CUV++2}
\vacave{L_{int} \: T_{UU}T_{VV} L_{int}}_c \nn
&& - \; \re \left[ \vacave{T_{UU} T_{VV} L_{int} L_{int}}_c  \right] \ .
\ea
Notice that the second term behaves like the second term in the
mean flux, see \reff{TVVI}:
it vanishes when integrated over $V'$ (or $U$).
Moreover, when applied to an accelerated mirror it leads to a
vanishing $C^{+/+}(U,V')$ (as the second term in  \reff{TVVI}
leads to a vanishing mean flux) when the coupling is constant
and when $U$ and $V'$ are both in the causal future of the mirror.
 
\reff{CUV++2} guarantees that $C^{+/+}$ is real and only depends on $F$
defined in \reff{FUV}:
\ba
C^{+/+}(U,V')\label{CUV++3}
= \left( F^*(U,V') +  F(U,V') \right)^2 
= 4 \left( \re [ F(U,V') ] \right)^2 \ .
\ea
Since it arises only  from pair creation amplitudes,
this guarantees that it is finite, as for \reff{CUV-+uag2}.
When the trajectory is inertial, \reff{FUVinert2} gives
\ba \label{CUV++inert}
C^{+/+}_{inert}(U,V') = \frac{g_0^2}{(4\pi)^2}
\left[ 
\di_U \left( \frac{g(U)}{(U-V')^2} \right)
+ \di_{V'} \left( \frac{g(V')}{(V'-U)^2} \right)
\right]^2 \ .
\ea
When $g(t)$ is constant, $C^{+/+}_{inert}$ vanishes for every couple of points $(U,V')$.
This is as it should be: inertial systems do not radiate when their 
coupling to the radiation field is constant.
Moreover, when $g(t)$ varies,  \reff{CUV++inert} is finite even in 
the coincidence image limit, that is $V' \to V_{cl}(U)$. In this limit, one gets 
\ba
\lim_{V'\to U}C^{+/+}_{inert}(U,V') \label{limitinert}
&=& \frac{g_0^2}{(24\pi)^2}  
\left[ \partial_U^3 g(U) \right]^2\ .
\ea
This equation should be added to Eq.($95$) in \cite{OPa}
which gives the mean flux emitted by this inertial mirror.

When applied to a uniformly accelerated mirror, 
\reff{CUV++1} depends on the sign of $U$ and $V'$.
By causality, in three of the four cases, $C^{+/+}$ can be expressed 
in terms of the correlation functions previously computed between $\scrym$ and $\scryp$.
The fourth case is the analog of the configurations studied in the inertial case, 
\reff{CUV++inert}.
We first discuss the three other cases.

Since the $V'>0$ part of $\scrypL$ is causally
disconnected from the trajectory, one has
\ba
C^{+/+}_{unif.acc.}(U<0,V'>0)= C^{+/-}_{unif.acc.}(U<0,V'>0) =0 \ ,
\ea
according to \reff{vanishCUV-+ua}, see Fig.$3$, and 
\ba \label{corrCUV-+uag2}
C^{+/+}_{unif.acc.}(U>0,V'>0)= C^{+/-}_{unif.acc.}(U>0,V'>0) \ ,
\ea
given in \reff{CUV-+uag2}, see Fig.$4$.
Moreover, the $U<0$ part $\scrypR$ is also disconnected from the trajectory.
Hence, one obtains
\ba \label{nonvanishCUV++}
C^{+/+}_{unif.acc.}(U<0,V'<0)= C^{+/-}_{unif.acc.}(V'<0,U<0) \ .
\ea
By direct evaluation, one gets
\ba
C^{+/-}_{unif.acc.}(v_L',u_R)= C^{+/-}_{unif.acc.}(u_L = v_L', v_R' = u_R)  \ ,
\label{++LR}
\ea
where $-a U=e^{-au_R}$ and $-a V'=e^{-av_L'}$
and where the r.h.s. is given by \reff{CUV-+uag2}. 
   
The last and most interesting correlations are encountered when the supports
of $T_{UU}$ an $T_{VV}$ are both in the future of the trajectory,
\ie for $U>0$ and $V'<0$.
In the Davies-Fulling model, the two-point function is identically zero since
the mirror is perfectly reflecting.
Instead, the Lagrangian model provides a non-vanishing result.
In terms of Rindler coordinates, Eqs.(\ref{FUVuag}) and (\ref{CUV++3}) gives
\ba \label{CUV++ua}
C^{+/+}_{unif.acc.}(u_L,v_L') = \frac{g_0^2}{(4\pi)^2}
\left[
\di_{u_L}  \left( \frac{g(u_L)}{\frac{4}{a^2}\sinh^2(\frac{a}{2}(u_L-v_L')} \right)
+ \di_{v_L'}  \left( \frac{g(v_L')}{\frac{4}{a^2}\sinh^2(\frac{a}{2}(v_L'-u_L)} \right)
\right]^2.
\ea
Eqs.(\ref{CUV++inert}) and (\ref{CUV++ua}) enjoy similar properties.
Indeed, $C^{+/+}_{unif.acc.}$ also vanishes when $g$ is constant
and when $g(\t)$ varies, it is finite in the coincidence image limit
\ba
\lim_{v_L'\to u_L}C^{+/+}_{unif.acc.}(u_L,v_L') 
= \frac{g_0^2}{(24\pi)^2} 
\left[ (\partial_{u_L}^3 - a^2 \partial_{u_L} ) g(u_L) \right]^2 \ .\label{limitacc}
\ea
\reff{CUV++ua} is an illustration of Grove's theorem \cite{Groveth,PhysRep}
which states that the fluxes emitted by a uniformly accelerated system 
behave like those emitted by the same system when it is inertial and 
in a thermal bath. 
Indeed, \reff{CUV++ua} follows from \reff{CUV++inert} 
by replacing Minkowski coordinates by Rindler ones 
and the vacuum two-point function by its thermal expression.
One has only transient effects when the switching function varies.
This is illustrated in Fig.$5$ where one sees 
that the two-point function possesses
two peaks localized in transients.
\begin{figure}[ht] 
\epsfxsize=5cm
\centerline{{\epsfbox{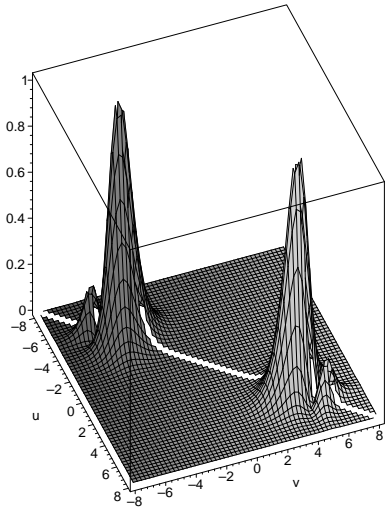}}}
\caption{$C^{+/+}_{unif.acc.}(u_L,v_L)$ when $g(\t)$ is given by \reff{g} 
and in arbitrary units. 
We choose $a=1$ and $\ln\eta=-4$. 
The two peaks are located for $u_L=v_L \simeq \pm (\ln2\eta)$,
when the coupling is switched on and off, see \reff{lapse}.}
\end{figure}

In brief, the lessons we obtained from the
analysis of the two-point function are the following,
see the two tables below.
For an accelerated mirror, the mean flux and the two-point function in causal 
contact, \reff{CUV++ua}, are concentrated in the transients.
Hence they both vanish in the limit $g(\t) \to const.$ 
and in the Davies-Fulling model. 
However, this is not the case for \reff{corrCUV-+uag2}.
Moreover, since the latter is a function of $u_L +v_R'$ (\ie $\t - \t'$)
when $g(\t)$ is a constant, this indicates a steady production of particles. 
To understand the origin of this discrepancy, 
we shall analyze in the next Section
matrix elements of $T_{\mu \nu}$ which contain more information
than the two-point function.
\begin{table}[!ht]
\begin{center}
\begin{tabular}{|c||c|c|}
\hline
$\left( \frac{16\pi}{a^2g_0} \right)^2 C^{+/-}_{unif.acc.}(u,v')$ 
& $U=-\inv{a}e^{-au_R}<0$ 
& $U=\inv{a}e^{au_L}>0$ \\
\hline
\hline
$V'=-\inv{a}e^{-av_L'}<0$ 
& $0$
& $\left[ 
\di_{u_L}  \left( \frac{g(u_L)}{\sinh^2(\frac{a}{2}(u_L-v_L')} \right)
\right]^2 $ \\
\hline
$V'=\inv{a}e^{av_R'}>0$ 
& $0$
& $\left[ 
\di_{u_L}  \left( \frac{g(u_L)}{\cosh^2(\frac{a}{2}(u_L+v_R')} \right)
\right]^2$
\\
\hline
\end{tabular}
\end{center}
\caption{The two-point functions between $\scrym$ and $\scryp$ 
for a uniformly accelerated mirror in $L$. They are of three different results.
$C^{+/-}$ identically vanishes for $U<0$, since this corresponds to causally disconnected 
outgoing fluxes. On the contrary, $C^{+/-}$ 
diverges in the coincidence image limit, for $V'=V_{cl}(U)$.
Finally,  $C^{+/-}$ is finite in the remaining case which corresponds 
to the correlations between a $U$ quantum and its $V$-partner, see Fig.$4$.}
\end{table}
\begin{table}[!ht]
\begin{center}
\begin{tabular}{|c||c|c|}
\hline
$\left( \frac{16\pi}{a^2g_0} \right)^2C^{+/+}_{unif.acc.}(u,v')$ 
& $U=-\inv{a}e^{-au_R}<0$ 
& $U=\inv{a}e^{au_L}>0$ \\
\hline
\hline
$V'=-\inv{a}e^{-av_L'}<0$ 
& $\left[ 
\di_{v_L'}  \left( \frac{g(v_L')}{\cosh^2(\frac{a}{2}(u_R+v_L')} \right)
\right]^2$
& $\left[ \di_{u_L}  \left( \frac{g(u_L)}{\sinh^2(\frac{a}{2}(u_L-v_L')} \right)
+ \left\{ u_L \leftrightarrow v_L' \right\}
\right]^2$ \\
\hline
$V'=\inv{a}e^{av_R'}>0$ 
& $0$
& $\left[ 
\di_{u_L}  \left( \frac{g(u_L)}{\cosh^2(\frac{a}{2}(u_L+v_R')} \right)
\right]^2$
\\
\hline
\end{tabular}
\end{center}
\caption{The two-point functions on $\scryp$ for a uniformly accelerated mirror in $L$.
$C^{+/+}$ vanishes for disconnected configurations $V'>0$ and $U<0$.
In the three other cases it is finite everywhere as it results
from pair creation amplitudes only.}
\end{table}

\section{Quantum correlations revealed by ``post-selection''}

To further analyze the correlations we now apply a second method.
Two different situations are considered.

In Sec. $3.1$, the selection of the final configurations on $\scryp$
(hereafter called ``post-selection'' to conform ourselves to the jargon) 
is imposed from the outset. This method is straightforward and leads to 
the answer we are searching, namely to identify why $C^{+/+}$ in \reff{CUV++ua}
vanishes whereas $C^{+/+}$ of \reff{corrCUV-+uag2} indicates a steady pair creation
rate.  

In Sec. $3.2$ we shall nevertheless present another
way to perform the post-selection:
we shall couple the radiation field to an additional quantum device
and then post-select the state of the latter. 
The justification of this second approach is that the resulting expression 
for the conditional value of $T_{\mu \nu}$ will 
establish a clear relation between the traditional
approach based on the two-point function and the unusual one
based on the conditional values. We shall see that the conditional value 
is a kind of point-split and smeared version 
of the two-point function previously analyzed.

\subsection{The conditional value of $T_{\mu \nu}$}

To compute the conditional value of an operator, 
one first chooses some final configurations 
on $\scryp$ (in our case, a wave packet representing a
particle emitted toward $\scrypR$). 
This defines a projector ${\mathbf P}$ in Fock space.
Then, in the Heisenberg picture, 
the corresponding conditional values of $T_{\mu\nu}$ reads
\ba 
\ave{T_{\mu \nu}}_{{\mathbf P}}
&\equiv& 
\frac{\vacave{T_{\mu \nu}\, {\mathbf P}}}
{\vacave{{\mathbf P}}} \label{Postselect} = 
\vacave{T_{\mu \nu}}+
\frac{\vacave{T_{\mu \nu}\, {\mathbf P}}_c}
{\vacave{{\mathbf P}}}\ .
\ea
For further details concerning the physical meaning 
of the (connected) matrix element of $T_{\mu\nu}$, we refer to \cite{MaPa}.


\subsubsection{The choice of the wave packet}

To obtain simple expressions for $\ave{T_{\mu\nu}}_{{\mathbf P}}$ in the case
of a uniformly accelerated mirror, 
we shall post-select a one-particle state described by
a superposition of ``Unruh'' \cite{Unruh} modes ${\tilde \varphi_{\la}^U}$.
We recall that the Unruh modes are eigenmodes with respect to the proper time $\t$
evaluated along the trajectory
and superpositions of positive frequency Minkowski modes, for further details 
see \cite{Unruh,OPa2}.
We choose to perform the post-selection on $\scrypR$, in causal 
contact with the mirror, \ie for $U > 0$, see Fig.$2$.
The selected state is thus of the form
\ba \label{wp}
\ket{\Psi}=
\Ie{\la} f^*(\la ; \lab , \sigma , \ub_L) \: \a^{U\; \dagger}_\la \vac^{U,part} \ ,
\ea
where $\vac^{U,part}$ is the vacuum with respect to $U$ particles 
and $\a^{U\; \dagger}_\la$ is the creation operator of a Unruh $U$-particle.
The function $f$ is normalized as follows
\ba \label{norm1}
\bra{\Psi}\Psi\rangle = \Ie{\la} \left| f(\la ; \lab , \sigma , \ub_L) \right|^2 = 1 \ . 
\ea

We have chosen to select this state for two reasons. 
First we want to detect a particle which is produced by the scattering on the mirror, 
\ie we want $\ket{\Psi}$ to be orthogonal to the initial state,
the Minkowski vacuum $\ket{0}$. 
This requirement excludes to work with Rindler quanta 
since they are present in this state.
The second reason is obvious, the stationarity of the scattering
is expressed in term of eigenmodes of $i \partial_\t = \la$
which is a Rindler frequency. 

It is also important to mention that $\ket{\Psi}$ 
does not fully specify the state on $\scryp$. 
Indeed, since we are interested in determining the partner of $\ket{\Psi}$,
we do not specify what is the state of particles emitted to $\scrypL$ 
nor the anti-particle states.
Therefore $\ket{\Psi}$ defines a projector 
${\mathbf P}^{\ub_L}_{\lab , \sigma}$
(in the sense that 
${\mathbf P}^{\ub_L}_{\lab , \sigma} {\mathbf P}^{\ub_L}_{\lab , \sigma} =
{\mathbf P}^{\ub_L}_{\lab , \sigma}$) 
which only determines the $U$-particle sector:
\ba \label{operatorP}
{\mathbf P}^{\ub_L}_{\lab , \sigma} \equiv 
\ket{\Psi}\bra{\Psi} 
\otimes {\mathbf 1}[\a^{V},\b^{U},\b^{V}] \ ,
\ea
since ${\mathbf 1}[\a^{V},\b^{U},\b^{V}]$ 
is the identity operator for the particle states on $\scrypL$ 
and the anti-particle states.

To obtain analytical expressions, 
we choose the function $f$ to be
\ba \label{wpf}
f(\la ; \lab , \sigma , \ub_L) 
\equiv  
\frac{e^{i\la \ub_L}}{\sqrt{4\pi\la(e^{2\pi\la/a}-1)}} 
\frac{e^{-(\la-\lab)^2/2\sigma^2}}{\sqrt{2\pi\sigma}} \times
\la \sinh^2(\pi\la/a) \times {\cal N}\left[ \lab,\sigma \right] \ .
\ea 
The first factor corresponds to  
a Unruh quantum centered around $u_L = \ub_L$
with a mean frequency given by $\lab$.  
The second factor has been chosen so as to obtain analytical expressions
for the conditional values and the third factor
ensures that $\ket{\Psi}$ is normalized according to \reff{norm1}. 
Since we want to detect the particle well localized around $\ub_L$,
\ie around $a\bar U = e^{a \ub_L} > 0$,   
$\lab$ must obey
\ba\label{condwp1}
\lab < 0 \; \; \and |\lab|/a \gg 1 \ .
\ea
The first condition arises from the fact that Unruh modes of negative 
Rindler frequency live mainly in the $L$ sector. The
second condition guarantees that the second peak of the wave packet
found around $U= -\bar U <0$ is negligible. We recall that 
wave packets built with Unruh modes possess two peaks. 
The relative weights of their norms is the thermal
factor $ e^{- 2\pi \vert \lab \vert /a}$ encoding the Unruh effect.

Before computing the conditional value of the flux, 
one inquires into the Minkowski frequency content of $\ket{\Psi}$. 
To this end, we compute the probability to find 
a one-particle state of Minkowski frequency $\om$
\ba
\bvac a^U_\om \, {\mathbf P}^{\ub_L}_{\lab , \sigma}  \, 
a^{U \dagger}_\om \vac =
\left| \Ie{\la} f^*(\la ; \lab , \sigma , \ub_L) \,
\gamma_{\lambda \omega} \right|^2 \ ,
\ea
where 
\ba 
\gamma_{\lambda \omega} &\equiv& \vacave{a^U_\om \: \a^{U \: \dagger}_\la} \nn
&=& \frac{\Gamma(i\la/a)}{\sqrt{\frac{a\pi}{\la\sinh(\pi\la/a)}}} 
(\frac{\om}{a})^{-i\la/a} \frac{1}{\sqrt{2\pi\om a}} \ .
\ea
In the limit $|\lab|/a \gg 1$
the stationary phase condition gives
\ba \label{saddle}
\la_{sp} = \lab - i\frac{\sigma^2}{a}\ln(\frac{- \om}{ \la_{sp} e^{-a\ub_L}}) \ . 
\ea
The norm of the overlap is maximum when the imaginary
part of $ \la_{sp}$ vanishes. Thus $\ket{\Psi}$ is made
of Minkowski frequencies centered around 
${\bar \om} = \vert \lab \vert e^{-a\ub_L}$.
The spread in $\om$ is $\sigma e^{-a\ub_L}$. 
When $\sigma < a $, our wave packet is thus well-peaked both
in Minkowski frequencies and in space-time.


\subsubsection{The conditional values of the fluxes}

To obtain the connected part of the conditional value, 
we first compute the denominator of the second term in \reff{Postselect}
which gives the probability to detect our chosen quantum.
To the second order in $g_0$, it is given by
$\vacave{L_{int}{\mathbf P}^{\ub_L}_{\lab , \sigma} L_{int} }$.
Since $\ket{\Psi}$ is expressed in terms of Unruh quanta, it is 
appropriate to re-express  
the transition amplitudes in terms of these 
rather than Minkowski quanta as in Eqs.(\ref{Bogo1}) and (\ref{Bogo2}).
The resulting amplitudes  will be noted $\tilde B_{\la\lap}$.
We have suppress the upper indices $U,V$
because they are all equal. 
Moreover they simply depend on the Rindler Fourier components of $g(\t)$:
\ba
\tilde B_{\la\lap} 
&\equiv& 
\vacave{\a_\la^U \: \b_\la^V \: iL_{int}} 
= \vacave{\a_\la^U \: \b_\la^U \: iL_{int}} \nn
&=&
i g_0
\frac{\la\lap}{\sqrt{\la(e^{2\pi\la/a}-1)\lap(e^{2\pi\lap/a}-1)}} 
\; g_{\la+\lap}^* \ . \label{amplU}
\ea
Then the probability reads
\ba \label{probability}
\vacave{L_{int}{\mathbf P}^{\ub_L}_{\lab , \sigma} L_{int} }
= 2 \int \! \!\int \! \!\Ie{\la} d\lap  d\mu \:
\tilde B_{\la\mu}^* \tilde B_{\lap\mu} f^*(\la) f(\lap) \ .
\ea
We must now verify that the matrix elements computed with $  \tilde B_{\lap \la}$ converge.
We remind the reader that when computing the {\em average} value of the fluxes, 
convergence was provided by the asymptotic decreasing of $g(\t)$, 
namely $g(\t) \to 0$ faster than $e^{- a |\t|}$, see \cite{OPa2}.
In the case of {\em conditional} values, a UV frequency cut-off
can be provided by $f(\la)$ which characterizes the post-selected wave packet.
When this is the case, one can safely consider the limit of constant coupling: $g(\t)=1$.
In this limit, we get
\ba \label{Blalagst}
\tilde B_{\la \lap} 
= - \frac{i g_0}{2}
\frac{\la}{\sinh(\pi\la/a)} 
\; \delta(\la+\lap) \ . 
\ea

When using $f$ given in \reff{wpf}, the Gaussian weight guarantees
that all expressions are well-defined in the UV. 
Moreover, \reff{wpf} leads to analytical expressions for the probability 
and the conditional values of the flux.
However, since these expressions are rather complicated,
we shall present only their behavior in the limit
$|\lab|/a \gg \sigma^2/a^2$ in addition to \reff{condwp1}
(This condition means that we work with well peaked wave packets in $\la$.).
In this limit, the probability reads
\ba
\label{prob}
\vacave{L_{int}{\mathbf P}^{\ub_L}_{\lab , \sigma} L_{int} }
= 2 g_0^2 \lab^2 \: e^{-2\pi|\lab|/a} \ ,
\ea
which is independent of the value of $\ub_L$, 
thereby indicating a steady production of particles
weighted by the thermal factor, 
as expected from the general analysis of Appendix $C$ in \cite{MaPa}.
(The corrections to \reff{prob} and the following equations are $O(\sigma/\lab)$). 

The connected part of the conditional values of the flux is
\be \label{twofluxes}
\ave{T_{\mu \nu}}_{{\mathbf P}\: c}
= \ave{T_{\mu \nu}}_{{\mathbf P}}^{p.s.} + \ave{T_{\mu \nu}}_{{\mathbf P}}^{partner} \ ,
\ee
where $\ave{T_{\mu \nu}}_{{\mathbf P}}^{p.s.}$ 
is the flux carried by the post-selected particle
and $\ave{T_{\mu \nu}}_{{\mathbf P}}^{partner}$,
that carried by its partner.
When ${\mathbf P}$ is given by \reff{operatorP}, 
$\ave{T_{\mu \nu}}_{{\mathbf P}}^{p.s.}$ is purely outgoing 
and consists in matrix elements of $\tilde a^U \tilde a^{U \: \dagger}$ and $ \tilde 
a^U \tilde b^U$. 
It possesses two maxima for $U=\bar U$ and $U=- \bar U$, 
which are related to the probability 
of measuring the position of the post-selected wave packet.
Moreover, as for Hawking quanta emerging from a black hole,
it is complex \cite{MaPa}.

The ``partner'' term of \reff{twofluxes} is more interesting.
To second order in $g_0$, it is given by 
\ba 
\ave{T_{\mu \nu}}_{{\mathbf P}}^{partner}
= \label{TPpartner0}
\frac{\vacave{L_{int}T_{\mu \nu}{\mathbf P}^{\ub_L}_{\lab , \sigma} L_{int}}}
{\vacave{L_{int}{\mathbf P}^{\ub_L}_{\lab , \sigma} L_{int} }} 
\ea
Unlike the $p.s.$ term, $\ave{T_{\mu \nu}}_{{\mathbf P}}^{partner}$
contains both $U$ and $V$ fluxes. Moreover, since the scattering 
amplitudes of \reff{amplU}
are identical for $U$ and $V$ modes,
one has $\ave{T_{UU}(U)}_{{\mathbf P}}^{partner}=\ave{T_{VV}(V=-U)}_
{{\mathbf P}}^{partner}$. Explicitly one gets
\ba 
\ave{T_{VV}}_{{\mathbf P}}^{partner}
= \label{TPpartner}
\frac{2 \disp \left|
\disp \int \!\! \Ie{\la} \! d\lap 
{\tilde B_{\la \lap}} f(\la)
\di_V {\tilde \varphi_{\lap}^V}
\right|^2}{\vacave{L_{int}{\mathbf P}^{\ub_L}_{\lab , \sigma} L_{int} }}  \ ,
\ea
where $\tilde \varphi_{\la}$ is the mode associated with the 
Unruh operator $\tilde a_\la$.  
This part of the conditional flux consists only in matrix elements of 
$\tilde b^V \tilde  b^{V \: \dagger}$.
Having post-selected a wave packet made with $\a$ only, 
this guarantees that the ``partner'' term is real.

In the right quadrant, for $V>0$, the Rindler flux is 
\ba
\label{TvRvR}
\ave{T_{vv}(v_R)}_{{\mathbf P}}^{partner}
= \frac{\sigma |\lab|}{\sqrt{\pi}}  
e^{-(\ub_L+v_R)^2\sigma^2} \ . 
\ea
In this, we find a behavior very similar to that of $C^{+/+}(\bar u_L, v_R)$
of \reff{corrCUV-+uag2}: $C^{+/+}$ also exhibits a constant maximum 
for $v_R=-\ub_L$, see Table $2$.
The width is here given by $1/\sigma$ instead of $1/a$
as in \reff{corrCUV-+uag2},
because of our choice of the window function $f(\la)$ in \reff{wpf}.
The similarity of $ \ave{T_{vv}(v_R)}_{{\mathbf P}}^{partner}$ 
when having post-selected a $U$-quantum at $\ub_L$,
with $C^{+/+}(\bar u_L, v_R)$ should cause no surprise: 
the first term of \reff{CUV++2} is dominant and $T_{UU}$ acts in it 
as ${\mathbf P}$ does it in \reff{TPpartner0}.

However
the correspondence with the two-point function is lost
when computing the conditional value in the left quadrant, for $V<0$. 
Whereas $C^{+/+}(u_L,v'_L)$ vanishes for $g(\t)=1$, we obtain
\ba
\label{TvLvL}
\ave{T_{vv}(v_L)}_{{\mathbf P}}^{partner}
= \frac{\sigma|\lab|}{\sqrt{\pi}}  
e^{-(\ub_L-v_L)^2\sigma^2}  e^{-2\pi|\lab|/a}\ ,
\ea
which is smaller than \reff{TvRvR} by the thermal factor $e^{- 2\pi |\bar \la|/a}$. 
The origin of this loss is as follows.
In \reff{CUV++1} the post-selection induced by $T_{uu}(u_L)$ is strictly confined in the
$L$ quadrant. Hence it is insensitive to the transients (located on $U=0^+$ in the
limit $g \to const.$) which contain
all the emitted particles. On the contrary, the post-selection carried by $\ket{\Psi}$ 
specifies that one Unruh quantum be present on $\scrypR$. This prescription is
sensitive to the particle content of the 
transients. In fact, {\it all} wave packets made with only positive Minkowski frequency modes
are sensitive to these transients, since there exists {\it no} such wave packet
which can vanish in a Rindler quadrant.\footnote{The first 
explanation of the compatibility of a null mean local flux 
with the readings of a particle detector was made in \cite{Grove}. Grove
proved that the detection of particles produced by a uniformly accelerating mirror 
occurs only if the detector is switched on in the causal future of the transients, see also
\cite{Unruh92} for a similar observation in a slightly different context.} 
Had we post-selected 
a superposition of $L$ Rindler quanta only, we would have found that 
$\ave{T_{vv}(v_L)}_{{\mathbf P}}^{partner}$ 
identically vanished, exactly like $C^{+/+}$ of \reff{CUV++ua}. 
The origin of this null result can be traced back to \reff{CUV++2}. 
When one post-selects only $L$ 
Rindler quanta, the contribution of the second (interfering) term 
cancels that of the first term.
Instead when imposing that a superposition of Minkowski quanta be 
found on $\scryp$, the second term vanishes since the post-selected 
state is orthogonal to Minkowski vacuum. 

We learned from this analysis that, being local in $U$,
$C^{+/+}(u_L,v'_L)$ is an extremely coherent object
whose vanishing results from fine tuned interferences.
The slightest modification of the scattering, \eg recoil effects\cite{Recmir,OPa4}
or switching off effects, 
would break these interferences.
This leads to a non-vanishing result 
whose content of Unruh quanta tells us that the pair creation process
is stationary.
In conclusion, the two-point correlation function vanishes because, on one hand,
it probes only locally the final configurations, and, on the other, 
the description of the scattering is too simplified.

When looking at the partner conditional flux
on $\scrypR$, one obtains the same expressions as in Eqs.(\ref{TvRvR}) and (\ref{TvLvL})
with $v \to u$
\ba
\label{TuRuR}
\ave{T_{uu}(u_R)}_{{\mathbf P}}^{partner}
&=& \frac{\sigma|\lab|}{\sqrt{\pi}}  
e^{-(\ub_L+u_R)^2\sigma^2} \\
\label{TuLuL}
\ave{T_{uu}(u_L)}_{{\mathbf P}}^{partner}
&=& \frac{\sigma|\lab|}{\sqrt{\pi}}  
e^{-(\ub_L-u_L)^2\sigma^2}  e^{-2\pi|\lab|/a}\ .
\ea From these, one sees that 
the $U$ partner of a $U$ post-selected wave packet
lies mainly on the other side of the horizon $U=0$, see Fig.$6$, 
and is distributed in a way which once more displays 
the stationarity of the process.

\begin{figure}[ht] 
\epsfxsize=7.5cm
\centerline{\epsfbox{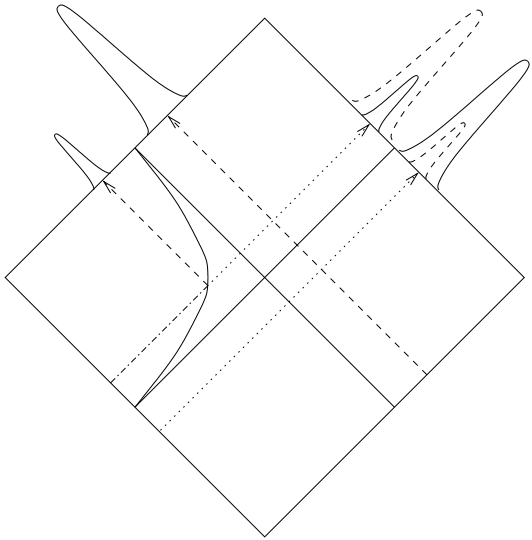}}
\caption{In this figure, the intensity profile of 
the post-selected wave packet is drawn as a dashed curve. 
The corresponding values of the partner conditional fluxes
are represented by plain curves. The $U/V$ symmetry of these
fluxes is manifest.
The dotted and dashed straight lines schematically represent the 
characteristics followed by the partner configurations.}
\end{figure}

In conclusion, we notice that the transition amplitude $B_{\la\lap}$ of \reff{Blalagst}
is unchanged if one now considers the scattering by a mirror moving in the right quadrant. 
Hence,  one would obtain exactly the same conditional fluxes.
To obtain expressions which depend on the side in which
the mirror lives, one should consider time dependent coupling.
This is the subject of Sec. $4$.
Before doing so, we shall consider another way to implement 
the post-selection which will reveal the relations between the
conditional fluxes and the two-point function.

\subsection{Post-selection by an additional quantum device}

Another way to implement the post-selection is 
to introduce an additional quantum system coupled to the field
on the right of the mirror.
In what follows, we shall use an inertial two-level atom\cite{BirrelDavies,PhysRep} 
positioned at $z= const$, on the right of the mirror.
The transitions of the two-level atom are 
described by the lowering operator $A(t)=e^{-imt} A$ and its hermitian 
conjugate. 
Here, $m$ is the energy gap between
its ground ($\ket{-}$) and excited state ($\ket{+}$).
One has $A\ket{-}=0$ and $A\ket{+}=\ket{-}$.

To make contact with Sec. $3.1$,
we couple the detector only to $U$ quanta. This is achieved by 
the following action
\ba \label{L_A}
L_A
= - f_0 \int \! dU f(U) 
\left(
\di_U \Phi^{U \; \dagger} A e^{-imU} 
+ \di_U \Phi^{U} A^\dagger e^{imU}
\right) \ .
\ea
Here, $f_0$ is the coupling constant 
and $f(t)$ is a real function
which governs how the interaction is turned off and on.

Instead of \reff{Postselect},
we consider the value of the energy flux $\ave{T_{VV}}_\Pig$
which is conditional to find the detector in its excited state 
at $t= +\infty$ when the initial state is $\vac \otimes \ket{-}$.
The post-selection is imposed by applying the projector 
$\Pig = \ket{+} \bra{+}$ 
at $t=+\infty$. 
To second order in $f_0$, in the Heisenberg representation
for the evolution governed by $L_{int}$, we obtain
\ba \label{TPi}
\ave{T_{VV}}_\Pig
&=& \vacave{ T_{VV}} +  \frac{\vacave{T_{VV} \, \tilde\Pig}_c}
{\vacave{\tilde\Pig}}  \ ,
\ea
where
\ba
\tilde\Pig = \bra{-} L_A \: \Pig \: L_A \ket{-} 
=  f_0^2 \: \int\!\!\Ie{U} \! dU' f(U) f(U') \, e^{-im(U-U')}
\: \di_U \Phi^\dagger \di_{U'} \Phi \ .
\ea
In the connected part of the conditional flux, 
the time ordering of $L_A$ with $L_{int}$
is such that $L_A$ can always be sent on the
future of the evolution operator $Te^{iL_{int}}$
since the detector is on the right of the mirror
and since it responds only to $U$ quanta.
In the interacting picture, 
the second term in \reff{TPi} is
\ba \label{TPi2}
\ave{T_{VV}}_{\Pig\: c}
=
\frac{ \disp\int\!\!\Ie{U} \! dU' f(U) f(U') \, e^{-im(U-U')} \,
\vacave{e^{-iL_{int}} \: \di_U \Phi^\dagger \di_{U'} \Phi \: \TVV \: e^{iL_{int}}}_c}
{(\vacave{e^{-iL_{int}}\: \tilde\Pig\: e^{iL_{int}}}/f_0^2)} \ .
\ea
This expression should be compared to the two-point function of \reff{CUV++1}
and to the former conditional flux, see \reff{TPpartner0}.
Two limits can be considered.
In the first limit, $f$ is localized in space-time, \ie $f(U)=\delta(U-U_0)$.
In this case, the numerator gives 
the two-point function
$C^{+/+}(U_0,V)$ of \reff{CUV++1}
whereas the denominator gives the mean value of $\TUU$.
In the second limit, the Fourier transform of $f e^{-im U}$ is
``local'' in the energy space,  \ie $\tilde f(\om) = \delta(\om - m)$.
In this case,
the particle detector is switched on for
all times and is sensitive only to Minkowski quanta of frequency $m$.
Then, up to an overall constant, 
the projector $\tilde \Pig$ reduces to the 
normal-ordered counting operator
${\mathbf P}_m = a^{U \: \dagger}_m a^U_m$.
When acting on one-particle Minkowski states,
it thus acts as the projector of \reff{operatorP}.

We have thus proved that the conditional value, \reff{TPi2}, 
generalizes the notion of correlation functions as
it interpolates from the local two-point function $\ave{T_{UU}T_{VV}}$ 
to the global correlation $\ave{a^{U \: \dagger}_m a^U_m T_{VV}}$ 
which relies on the notion of particle. The intermediate cases
correspond to smeared and point-split expressions,
see \cite{Hu2,Ford} for similar considerations on smeared correlation functions.

\section{Scattering by two accelerated mirrors}

In this Section,
we study the scattering by two accelerating mirrors
which follow symmetrical trajectories in the $R$ and $L$ quadrants (see Fig.$9$).
In Sec. $4.1$, using the (unregulated) Davies-Fulling model, we show that
the Bogoliubov coefficients encoding pair creation identically vanish. 
Their vanishing arises from
perfect interferences, in very much the same way of the perfect interferometer 
of Gerlach\cite{Gerlach}.
In Sec. $4.2$, we prove that these perfect interferences are an artifact of the
unregulated description in which the coupling of the mirror to 
the radiation field is strictly constant.
In Sec. $4.3$, we show how this singular
regime can be approached (but never fully realized if one insists
on keeping regularity)
by fine-tuning the coupling constant.

\subsection{The Davies-Fulling description}

In the Davies-Fulling model,
the mode scattered by the two mirrors is given by 
\ba
\varphi^{scat}_\om (U)= - \frac{e^{-i\om V_{cl}(U)}}{\sqrt{4\pi\om}} \ , 
\ea
for {\it all U}. 
Indeed, the peculiarity of two symmetrical uniformly accelerated trajectories 
is that $V_{cl}(U)= - 1/a^2 U$ is valid for all values of $U$ (and $V$),
as in the case of a single mirror which originates from $i^-$ and ends 
in $i^+$. If only one accelerated mirror was present,
the support of $V_{cl}(U)= - 1/a^2 U$ would have been restricted to half the real axis. 

The Bogoliubov coefficient ${}^{L+R}\bt^{UV \: *}_{\om\omp} $
encoding pair creation is given,
as usual, by the overlap between $\varphi^{scat}_\om$
with an $out$-mode of frequency $\omp$, see \reff{bUV}. Thus one has
\ba
{}^{L+R}\bt^{UV \: *}_{\om\omp} 
&=& {}^L\bt^{UV\: *}_{\om\omp} + {}^R\bt^{UV\: *}_{\om\omp} \ ,
\ea
where ${}^L\bt$ (${}^R\bt$) is the Bogoliubov coefficient one would obtain
 when considering only one mirror.
Moreover since  \cite{BirrelDavies} 
\be
{}^R\bt^{UV \: *}_{\om\omp} = \frac{i}{\pi a} K_1(2\sqrt{\om\omp}/a) = - 
\; \; {}^L\!\bt^{UV \: *}_{\om\omp}\ ,
\ee
the total Bogoliubov coefficient
${}^{L+R}\bt^{UV}_{\om\omp}$
vanishes for all values of $\om$ and $\omp$ !!! 
Thus the total energy emitted 
\ba
\label{DFH}
\ave{H_M^V} =
\Io{\om} \om \Io{\omp} 
\left| 
{}^{L+R}\bt^{UV \: *}_{\om\omp} 
\right|^2
\ea
vanishes as well since it is given in terms of the square of 
${}^L\bt^{UV } + {}^R\bt^{UV }$
and not in terms of the sum of their squares. 
Therefore,  $\ave{H_M^V}$ vanishes 
because of the perfectly destructive interferences
between the scattering amplitudes.

In brief, in this description, no pair is created,
\ie the two mirrors have no effect on the vacuum configurations,
exactly like an inertial mirror. This cancellation is directly related to 
what Gerlach called a perfect interferometer,
see Eq.(125) in \cite{Gerlach}. 
It is also related to the canceling effect found by Yi\cite{Yi} 
when considering the asymptotic radiation emitted by 
two accelerated (charged) black holes. 

It is important to verify that this result is not due to the fact that
the mirrors are perfectly reflecting. In fact, it is
also obtained when using partially transmitting mirrors
in constant interaction with the radiation field.
This is easily verified by using the interacting model.
In this case, each mirror is coupled to the field 
by a Lagrangian given by \reff{Hint}
with coupling parameter 
$g_0^L$ ($g_0^R$) and switching function $g^L(\t)$ ($g^L(\t)$).
The total Lagrangian is the sum  
$L_{int}=L^L+L^R$. Thus, using \reff{Bogo2}, to first order in $g_0^L$ and $g_0^R$,  
the amplitude of spontaneous pair creation
${}^{L+R}B_{\om \omp}^{ij}= \vacave{a_\om^i b_{\omp}^j \: iL_{int}}$ 
identically vanishes 
when\footnote{Had we taken the current $J=\Phi^\dagger i\didi{\tau} \Phi$
instead of 
$\di{\tau} \Phi^\dagger \di{\tau} \Phi +  \di{\tau} \Phi \di{\tau} \Phi^\dagger$
in \reff{Hint}, the condition would have been $g_0^R= g_0^L$,
thereby recovering the correspondence between the Lagrangian model 
and the Davies-Fulling one.} 
$g_0^R= - g_0^L$ and when $g^R(\t)=g^L(\t)= constant$.
Hence, perfect interferences do not follow from perfect reflection.
As we shall now prove,
they directly follow from the fact the coupling is constant.

\subsection{The regulated description}

In \cite{OPa2}, we proved that the scattering by a uniformly accelerated mirror
is regular only if its coupling to the radiation field decreases faster
than $e^{-a|\t|}$. 
(If this condition is not fulfilled, the expectation values of observables are ill-defined, 
\ie the result might depend on the order in which the integrations are performed.) 

We shall now prove that by using {\it any} regular description 
of the scattering, the interferences leading to the 
vanishing of the ${}^{L+R}\beta$ coefficients are {\it inevitably} lost.
To this end, it is convenient to work in a ``mixed'' representation, 
\ie with one quantum characterized by a Minkowski frequency and its partner by
a Unruh quantum with a given Rindler frequency, 
see Sec. $2.3$ in \cite{OPa2} for detailed expressions.
In this case, when using $g(\t)$ given in \reff{g},
the spontaneous pair creation amplitude is
\ba
{}^{R,L}B_{\om \la}^{VV}
&\equiv& \vacave{a_\om^V \: \b_\la^U \: e^{iL^{R,L}}}_c \nn
&=& \frac{ig_0}{\pi a} \; \sqrt{\frac{\om\la}{1-e^{-2\pi\la/a}}} \;
(1-is\om/a\eta)^{-(1+i\la/a)/2} \; 
K_{1+i\la/a}(2\eta\sqrt{1-is\om/a\eta})\, 
\ea
where $s=\pm 1$ for $R$ and $L$ respectively.
One also finds that ${}^{R,L}B_{\om \la}^{VU}= {}^{R,L}B_{\om \la}^{VV}$.
We thus see that the regulator $\eta \ll 1$, 
which is needed to obtain well-defined expressions,
enters differently in the amplitudes of the $R$ and $L$ mirrors. 
In fact, this is necessary for causality to be respected. 
Hence, it is quite conceivable that the former canceling effects
will be lost when $\om$ approaches the UV scale $a/\eta$.

Let us verify this numerically by studying the integrand of the Minkowski energy:
\ba \label{Hcoherent}
\ave{H_M^V}^{L+R} = 4 \Io{\om} \om \Ie{\la} 
\left| {}^{L}B_{\om \la}^{VV} + {}^{R}B_{\om \la}^{VV}\right |^2 
= \Io{\om} \Ie{\la} h_M^{L+R}(\om, \la; \eta) \ .
\ea

When $g_0^R= -g_0^L$ and when $g^R(\tau)= g^L(\tau)$ is given by \reff{g},
the integrand $h_M^{L+R}$ is shown in Fig.$8$.
\begin{figure}[ht] \label{wBR_lneta_lnomega}
\epsfxsize=6.5cm
\epsfysize=5.5cm
\psfrag{lnomega}{${\ln\om}$}
\psfrag{lneta}{${\ln\eta}$}
\centerline{\epsfbox{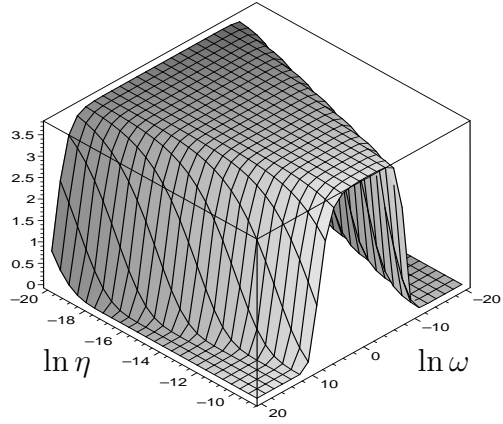}}
\caption{The integrand for one mirror $h_M^{L}(\om,\la=0.15 a;\eta)$ 
in terms of $\ln\om$ and $\ln\eta$ and in arbitrary units. 
One sees clearly that the contribution of Minkowski modes
exhibits a ``plateau'' of constant height which is 
limited by the cut-offs $\ln\om = \pm \ln\eta$. 
All the modes whose frequency belongs to this interval 
participate to the emitted flux.}
\end{figure}
\begin{figure}[ht] \label{wBR-BL2_lneta_lnomega}
\epsfxsize=6.5cm
\epsfysize=5.5cm
\psfrag{lnomega}{${\ln\om}$}
\psfrag{lneta}{${\ln\eta}$}
\centerline{\epsfbox{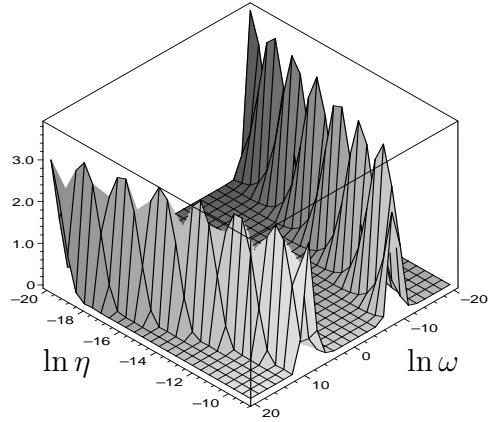}}
\caption{The integrand for the two mirrors $h_M^{L+R}(\om,\la=0.15 a;\eta)$ 
in terms of $\ln\om$ and $\ln\eta$ and in the same arbitrary units. 
Unlike for a single mirror, the ``inside'' modes with 
$ a \eta  \ll \om \ll a /\eta$ no longer contribute because
of the destructive interferences.
However, these interferences are lost when reaching the
transient frequencies $\om=a\eta$ and $\om=a/\eta$, thereby giving rise
to a positive integrand.}
\end{figure}
One clearly sees how $g(\t)$ plays its role.
The transients associated by the switching function are different
for the right and left mirrors. 
Hence the sum of amplitudes no longer vanishes for frequencies 
close to the frequency cut-offs $a\eta$ and $a/\eta$.
Instead, all the modes within the plateau,
\ie when the interaction is almost constant, interfere destructively,
as in the Davies-Fulling model. 
Hence, they do not participate to the total energy.

There is another simple way to prove that the two mirrors
cannot interfere destructively. It follows from causality.
In fact, since the two mirrors are causally disconnected, 
the mean flux obeys
\be
\ave{\TVV(V)}^{L+R} =  \ave{\TVV(V)}^{L} + \ave{\TVV(V)}^{R} \ ,
\ee
since $\ave{\TVV}^{L(R)}$ vanishes identically in $R(L)$.
Hence 
\ba
\ave{H^V_M}^{L+R} 
&=&  \ave{H^V_M}^{L} +  \ave{H^V_M}^{R} \nn
&=& 4 \Io{\om} \om \Ie{\la} 
\left(
\left| {}^{L}B_{\om \la}^{VV} \right |^2 
+ \left| 
{}^{R}B_{\om \la}^{VV} \right |^2 
\right) \ . \label{Hincoherent}
\ea
This tells us that whatever are the couplings 
between the two mirrors and the radiation field,
when causality is respected,
the total energy emitted by the mirrors can be added incoherently, 
\ie, {\em unlike} what was found in the Davies-Fulling model in \reff{DFH}.
Therefore, when causality is respected, Eqs.(\ref{Hcoherent}) and (\ref{Hincoherent}) imply
\ba
\Io{\om} \om \Ie{\la} \re \left\{
{}^{R}B_{\om \la}^{VV} \; {}^{L}B_{\om \la}^{VV \: *} \right\} 
\equiv 0 \ .
\ea

To complete the analysis we now determine to what extend 
two accelerated mirrors can constitute a perfect interferometer. 
That is : What is the most interfering situation ?

\subsection{A perfect interferometer ?}

To answer this question, we use the interacting model 
and we parametrize the coupling constant in Fourier transform:
$g^L(\t)=\Ie{\la} g^L_\la \: e^{-i\la\t}$ and 
$g^R(\t)=\Ie{\la} g^R_\la \: e^{-i\la\t}$. 

Instead of studying the average value of $H$ or $T_{VV}$,
it is simpler to compute the probability to receive one Unruh quantum on $\scryp$.
The two accelerating mirrors would provide a perfectly interfering device
if this probability vanishes. 
To compute it we use the projector ${\mathbf P}_{\lab}$
associated with the detection of a Unruh quantum on $\scrypR$, 
\ie with 
\be
f(\la) = \delta(\la-\lab) \frac{e^{i\la\ub_L}}{\sqrt{4\pi\la(e^{2\pi\la/a}-1)}}
\ee
in \reff{wpf}.
To the second order in the $g_0$'s, from \reff{probability}, we get 
\ba \label{2MP}
\ave{{\mathbf P}_{\lab}}^{L+R}
&=& \frac{n_{\lab}^2}{2\pi} \Ie{\la} \la n_\la
\left| g_0^L g^{L \: *}_{\la+\lab}
+ e^{\pi(\lab+\la)/a} g_0^R g^R_{\la+\lab} \right|^2 \ .
\ea
The above expression vanishes if 
\ba
g_0^L g^{L \: *}_\la = - e^{\pi\la/a} \: g_0^R g^R_{\la} \ , 
\ea
for all $\la$.
The only solution for (real) coupling functions 
is given by time-independent opposite real numbers :
$g^L(\t) = g^R(\t) = 1$ and $g^L_0 = - g_0^R$.
Therefore, we find that one can form a perfect interferometer
{\em if and only if} one considers two mirrors constantly in interaction 
with the radiation. However, the necessary condition 
to obtain regular Minkowski expressions is then violated. 

So let us minimize the probability $\ave{{\mathbf P}_{\lab}}^{L+R}$
by insisting that one has regular expressions, 
\ie that both $g_R$ and $g_L$ decrease faster that $e^{-a|\t|}$.
For instance, take Gaussian switching functions 
for each mirror with {\em a priori} different width and center:
\ba
g^R(\t) = e^{-(\t-\t_R)^2 / 2 T_R^2} \; \; \and
g^L(\t) = e^{-(\t-\t_L)^2 / 2 T_L^2} \ .
\ea
For simplicity and from our previous analysis,
we suppose that $g^R_0 = - g_0^L$.
We also work 
in the double limit of rare events $|\lab|/a \gg 1$ 
and long couplings $a T_{L,R} \gg 1$.
Then the ratio of the probabilities when having one or two mirrors is
\ba \label{interf}
\frac{\ave{{\mathbf P}_{\lab}}^{L+R}}{\ave{{\mathbf P}_{\lab}}^{L}}
\propto \inv{a^2T^2} (1+\chi_1 (\t_R+\t_L)^2a^2 + \chi_2 |T_R-T_L|/T_L) \ ,
\ea
where $\chi_1$ and $\chi_2$ are positive factors 
of order unity and $T=(T_R+T_L)/2$.
Therefore, the minimization of the probability is reached for switching functions:
\begin{itemize}
\item which possess the same lapse $T_R=T_L=T$
\item and which are centered symmetrically: $\t_R+\t_L=0$ (see Fig.$9$). 
\end{itemize}
When $T\to + \infty$, we get 
$\ave{{\mathbf P}_{\lab}}^{R+L} / \ave{{\mathbf P}_{\lab}}^{L} \to 0$,
thereby approaching the perfect interferometer behavior.
However, in this case,
the Minkowski energy will diverge like $e^{aT}$.
\begin{figure}[ht] 
\epsfxsize=6cm
\centerline{{\epsfbox{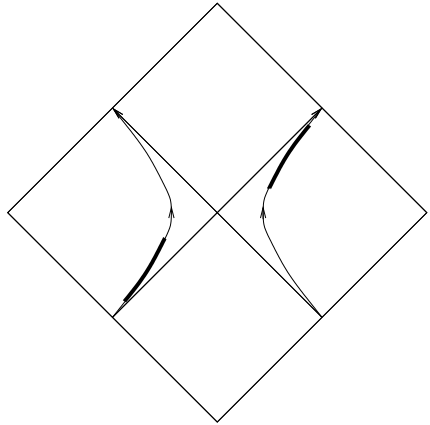}}}
\caption{In this diagram, we show the trajectories
followed by the two mirrors.
We represent an example of a fine-tuned device
by drawing thicker lines when the interactions are switched on.
The centers of the thicker parts are symmetrical with respect to $U=V=0$ 
and the lapses are identical.}
\end{figure}

\section*{Conclusions}

We recall the main results derived in this paper. 
We first study the quantum correlations within the fluxes 
emitted by  a uniformly accelerated mirror.
The results, which are summarized in Tables $1$ and $2$, 
reveal how the original correlations of the vacuum 
are scattered by the mirror. However, this
analysis is partial in that the particle content
of the fluxes is not disentangled when probing the final state by 
local operator. This is particularly clear for
$C^{+/+}$ of \reff{CUV++ua} which vanishes in the limit $g \to const$. 

To complete the analysis, we then compute the conditional flux
which is correlated to the detection of a given outgoing particle.
These conditional fluxes are rather similar to the corresponding 
two-point functions.
However, they differ in one respect:
the second term of \reff{CUV++2} never contributes to the conditional value 
whereas it does for the two-point function. 
In fact, it is at the origin of the vanishing of $C^{+/+}$ of \reff{CUV++ua}.
In addition, once its contribution is suppressed, either by a post-selection
or by taking into account recoil effects, one finds that the scattering
by a uniformly accelerated mirror leads to a steady production of 
pairs of quanta. 

To further relate the usual correlation functions to the 
conditional values of the fluxes, we present another way
to perform the post-selection so as to be able to recover the
formers as limiting cases of the latter.
We believe that this is quite important for the following reasons.
First, when studying quantum field theory in a curved space-time\cite{BirrelDavies},
one looses the notion of particle which exists
in Minkowski space-time. Therefore, in the absence of 
a unique definition of particle, it has been claimed that the
only meaningful quantities are expectation values of local operators.
We find this claim too restrictive in that not only mean quantities 
have physical meaning in quantum settings: 
To ask what happened when a particle detector
did click, or did not click, are perfectly legitimate questions.
It is therefore of importance to establish an explicit relationship
between the local and the particle descriptions. 
This has been here achieved by introducing a particle detector, 
computing the conditional value of the flux, and adjusting
the window function both in space-time and in the energy space, 
so as to generate a spectrum of matrix elements which reduce,
in limiting cases, to conventional expectation values of local operators 
and global operators.

Secondly, the correlations of local operators are singular in the
coincidence point limit and near a black 
hole horizon\cite{Jacobson91+Jacobson00,tHooft,Unruh95,BMPS95,MaPa,Parentani} 
or in inflationary cosmology\cite{MBrand+N,NP}.
It is still unclear at present how to handle these divergences
when considering the gravitational back-reaction 
associated with the fluctuations of $T_{\mu\nu}$.
It is thus also of importance to deal
from the outset with point-split and smeared generalization of 
(ultra-local) correlation functions. Such a generalization is naturally obtained 
by considering the conditional fluxes as computed in Section 3.2.

Finally, to further illustrate the need of using regularized transition amplitudes, 
we studied the scattering  by two mirrors
which follow symmetrical trajectories.
This example is particularly interesting since it leads to 
incoherent results when using the Davis-Fulling model.
Instead, when using regularized amplitudes, 
the apparently paradoxical results are all resolved.


\appendix

\section*{Appendix : Uniformly accelerated mirrors}

Uniform acceleration means that (up to a 2D translation)
\ba
t_{cl}^2(\t) - z_{cl}^2(\t) = - \inv{a^2} \ .
\ea
This equation defines two causally disconnected trajectories, 
lying respectively in $R$ and $L$, the right and left Rindler wedges.
In Sec. $2$ and $3$, we consider the scattering on the right 
of an accelerated mirror in the left quadrant. In null coordinates, its trajectory 
is given by
\ba 
V_{cl}(\t) 
= - \frac{1}{a} e^{-a\t} = - 1/ a^2U_{cl}(\t)   \ ,
\ea
(we work with $dt_{cl}/d\t>0$).

In the self-interacting model, as shown in \cite{OPa2},
any function $g(\t)$ which decreases faster than $e^{-a|\t|}$ 
is sufficient to obtain regular transition amplitudes.
A convenient choice is provided by
\ba \label{g}
g(\t) \equiv e^{-\disp 2 \eta \disp \cosh(a\t)} 
= e^{-\disp \eta  (aV_{cl}(\t)+1/aV_{cl}(\t))}
\ ,
\ea
where $0<\eta \ll 1$ is a dimensionless parameter.
This function possesses 
a plateau of height $1$ centered around $\t=0$
whose duration is given by 
\ba 2T \simeq \frac{2 | \ln(2\eta) |}{a} \ .\label{lapse} \ea
The slope of the switching on and off is independent of $\eta$ and 
proportional to $a$.
The tail decreases as $ e^{-\eta e^{a|\t|}}$, thus faster than
the required  $e^{-a|\t|}$.

Using \reff{g}, we obtained analytical expressions
for the scattering amplitudes $A^{ij \: *}_{\om\omp}$ and $B^{ij}_{\om\omp}$:
\ba \label{regABM}
A^{VV \; *}_{\om \omp} 
= \delta(\om-\omp) 
- \frac{4 i g_0}{\pi} \frac{\sqrt{\vert\om \omp\vert}}{a} \frac{\eta^2}{X^2} K_2(X) \ ,
\ea
where $X=\disp 2 \eta \sqrt{1 - i (\om - \omp)/a\eta }$,
and where $K_2$ is a modified Bessel function, see
Appendix $A$ in \cite{OPa2} and \cite{Abramovitz}, and 
\ba
A^{VU \; *}_{\om \omp} 
= - \frac{i g_0}{\pi} \frac{\sqrt{\vert\om \omp\vert}}{a} K_0(Y) \ ,
\ea
where $Y=2\sqrt{(\om/a +i \eta)(- \omp/a - i \eta)}$. 
The well-defined analytical properties 
of  $ A_{\om \omp}$ allows to obtain the
pair creation amplitudes by crossing symmetry:
\ba 
B^{ij }_{\om \omp} = - A^{ij \: *}_{\om,\om''=\omp e^{-i\pi}} \ . 
\label{crossym} 
\ea
Unlike the overlaps of Eqs.(\ref{aUV}) and (\ref{bUV}) evaluated with the Davies-Fulling model, 
see Eqs.($19$) of \cite{OPa2} for explicit expressions, 
these amplitudes are regular and well-defined 
in the complex plane of $\om$ and $\omp$.


\end{document}